\documentclass[twocolumn,prb, aps,amsmath,showpacs,floatfix]{revtex4}
\usepackage{graphicx}

\graphicspath{{figures/}}

\begin{document}

\title{Design and Performance of a Practical Variable-Temperature Scanning Tunneling Potentiometry System}
\author{M. Rozler}
\author{M. R. Beasley}
\email[Electronic address: ]{beasley@stanford.edu}
\affiliation{Geballe Laboratory for Advanced Materials, Stanford University, Stanford, California 94305}

\date{\today}

\begin{abstract}
We have constructed a scanning tunneling potentiometry system capable of simultaneously mapping the transport-related electrochemical potential of a biased sample along with its surface topography. Combining a novel sample biasing technique with a continuous current-nulling feedback scheme pushes the noise performance of the measurement to its fundamental limit - the Johnson noise of the STM tunnel junction. The resulting 130~nV voltage sensitivity allows us to spatially resolve local potentials at scales down to 2~nm, while maintaining angstrom scale STM imaging, all at scan sizes of up to 15~$\mu$m. A mm-range two-dimensional coarse positioning stage and the ability to operate from liquid helium to room temperature with a fast turn-around time greatly expand the versatility of the instrument. By performing studies of several model systems, we discuss the implications of various types of surface morphology for potentiometric measurements.
\end{abstract}

%\pacs{74.78.Bz, 74.25.Sv, 74.25.Qt, 07.79.-v}

\maketitle

\section{\label{intro}Introduction}

Scanning tunneling potentiometry (STP) is a technique for measuring electro-chemical potential differences over very small distances using a scanning tunneling microscope (STM).  It was first demonstrated by Muralt and Pohl in 1986\cite{muralt}, just four years after the invention of the STM itself. In this measurement, an STM tip is used as a moving voltage probe to measure the spatially varying electro-chemical potential generated in response to a bias applied across the sample.  Unlike most atomic-scale scanning probes, the information provided is not limited to the local surface, but reveals behavior over the length scales characteristic of the transport processes governing the response of the sample. Topographical information is also obtained, allowing the correlation of transport properties with structural information.  The technique is similar to the use of a conducting tip contact AFM to measure the electro-chemical potential\cite{contactafm} but can achieve very much greater spatial resolution - readily on the nanometer length scale -- as we demonstrate in this paper.

Subsequent to the work of Muralt and Pohl, several groups demonstrated variations and improvements on the original approach, and in some cases made simple physical measurements.  Perhaps the most successful example is the work of Briner \textit{et al.}\cite{feenstra} , in which potentials of the residual resistivity dipoles around individual topographical features were observed.  But to date, the separate improvements demonstrated in these various works have not been combined in a practical STP instrument suitable for general nanoscale transport measurements, and certainly not over a temperature range from room temperature down to liquid helium temperatures. Notably, the system of Muralt and Pohl worked over large areas, but due to the DC nature of that measurement, and the majority of others that followed\cite{feenstra, kirtley,kent, besold, schneider,ramaswamy, grevin}, the noise and spatial resolution were orders of magnitude from the fundamental limits of STP. By using AC techniques, Pelz and Koch reached the fundamental noise limits but only over very small areas\cite{pelzlownoise}.  Moreover, to date no attempts have been made to deal with problems of contact resistance variation, as they are addressed by common four-point conventional transport measurements, and only one group reported extending STP to lower temperatures, down to 77~K\cite{kent}. 

In addition, the early measurements were very difficult to reliably interpret, due to suspected tip jumping - discontinuous change in the tunneling position on the tip caused by finite tip size on a rough surface\cite{kent, pelzkoch}. The large potential jumps, observed by many researchers, have been typically attributed to random resistor networks in the presence of high resistance grain boundaries\cite{kirtley,schneider,ramaswamy, grevin}, or, in one case, thickness change of the film\cite{besold}, as well as the tip-jumping artifacts.

In this paper, we present the design and performance of a practical STP system capable of measurements from room temperature down to 4~K with high sensitivity and spatial resolution.  We also confirm definitively the existence of voltage steps due to tip jumping but at the same time show that they are not a fatal limitation to the application of the technique.  The instrument we have built has demonstrated voltage sensitivity limited by Johnson noise of the tunneling resistance, which, combined with low thermal drifts, allows spatial resolution of the potential down to 2~nm at scan ranges up to 15~$\mu$m.  

This paper is organized as follows.  In Section~\ref{design}, we review how STP works.  After that, we describe the physical components of the system: the scanning tunneling microscope and cryogenics. In Section~\ref{nonhardware}, we present the electronic instrumentation and the measurement protocol used to generate quantitative topographic (STM) and potentiometric (STP) images. Section~\ref{performance} addresses the performance of some of the key elements of the system. In Section~\ref{initialmeas}, we present initial measurements that allow us to document the technical performance of the complete instrument.  In Section~\ref{rough}, we discuss data on some model systems that demonstrate the tip jumping phenomenon and that are of physical interest for their own sake. Finally in Section~\ref{improve}, we briefly describe some ideas that could help further expand the capabilities of our system.

\section{\label{design}Design of the instrument}

\subsection{\label{principles}Principles of operation}
\begin{figure}[h]
\includegraphics*[width=.4 \textwidth]{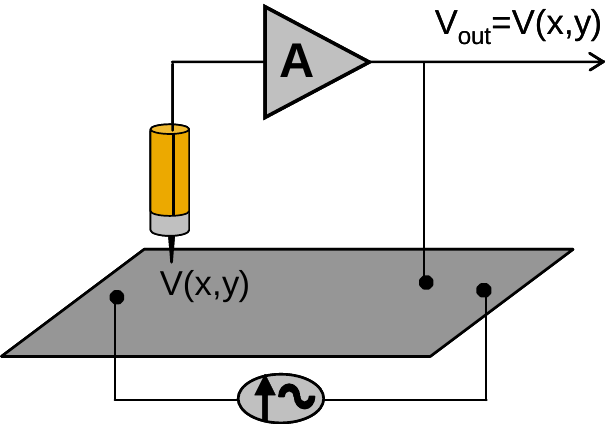}
\caption{Principle of operation of STP. Spatially dependent electrochemical potential is established by an AC current along the sample.
Tip-sample current signal is fed back to the sample; local sample potential is given by voltage required to zero the AC tip-sample current.  }
\label{fig:stpfig}
\end{figure}

The basic operation of STP is illustrated schematically in Figure~\ref{fig:stpfig}.  As in most transport measurements, a current bias (AC in our case) along the sample establishes a spatially dependent AC electro-chemical potential. AC biasing is used to avoid thermal emfs and 1/f noise. A scanning tunneling microscope is used alternately to perform STM (topography) and STP (potentiometry).  The two modes of operation are complementary in that in STM the DC tip-sample current is held constant at a preset value by feeding back on the vertical position of the voltage biased tip, whereas in STP, the AC component of the tip-sample current is kept at zero by feeding back an offset AC voltage on the entire sample so as to null the AC voltage across the junction. The AC offset that needs to be added to the sample in order to force the tip current to zero is a direct measure of the AC electrochemical potential under the tip.

The fact that in potentiometry there is no voltage drop across the tunnel junction has profound consequences for STP noise performance.  With the tip-sample voltage brought to zero, tunnel resistance fluctuations caused by tip vibrations of the microscope do not result in any electronic noise. The fundamental voltage sensitivity of an STP is thus set by Johnson noise in the junction resistance and is expected (and found) to be very much smaller than the sensitivities typically encountered in other STM-based measurements, such as  STS (scanning tunneling spectroscopy). The Johnson noise is 1.3 $\mu$V/$\sqrt{Hz}$ at room temperature for a typical 100~M$\Omega$ junction resistance; at 4.2~K, using a lower tip resistance of 1~M$\Omega$ , which lies at the low end of values possible for stable STM operation, it drops to 13~nV/$\sqrt{Hz}$, approaching the range normally associated with a standard fixed-contact lock-in measurement.  The practical measurement protocol used to switch conveniently and stably between STM and STP modes of operation are discussed in detail in Section ~\ref{protocol}. 
\subsection{\label{stm}Scanning Tunneling Microscope}
\begin{figure}[h]
\includegraphics*[width=.5 \textwidth]{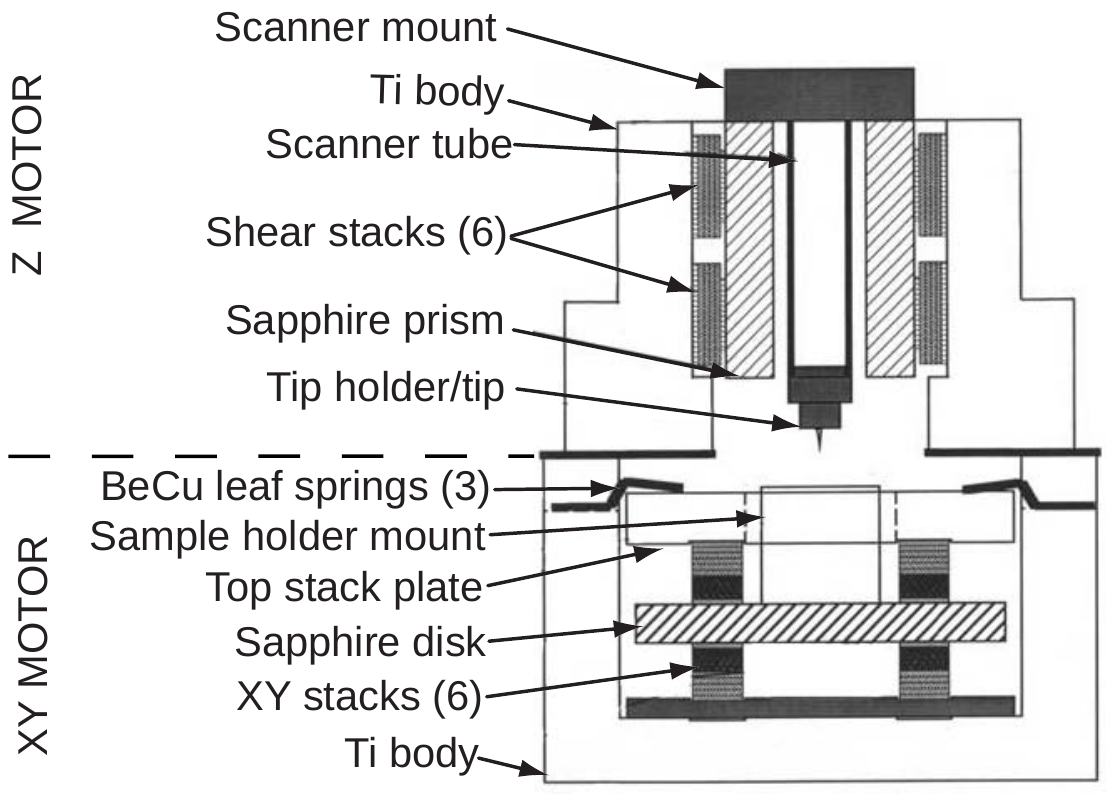}
\caption{Design of the STM. Coarse motion in both motors is achieved by sequential actuation of shear piezo stacks that are pressed against polished sapphire surfaces by BeCu leafsprings. In the Z motor, a sapphire prism holding a standard STM piezo tube (used for scanning) is translated vertically towards or away from the sample; in the XY motor a sapphire plate can be coarsely positioned laterally. }
\label{fig:stmpic}
\end{figure}

The main component of an STP system is a standard STM with several additional sample leads\cite{thesis}.  For the system described here, a vertically-oriented Pan-type microscope\cite{pan} was used. The Pan design was chosen for its reliable coarse motion, mechanical rigidity, and symmetric construction, which reduces thermal drifts.  In this design, a PZT tube scanner holding the STM tip is mounted on a polished triangular prism, pressed by means of a leaf spring against six shear piezo stacks (two on each face of the prism). Stick-slip translation is achieved by a step-function actuation of the stacks in a sequential fashion, followed by a simultaneous return of all stacks to the original position using a smoother waveform.  During individual stack steps, the prism is held in place by the five stationary stacks, while the actuated one slides along the prism's surface.  The prism is then moved relative to the microscope body (and the sample) during the simultaneous step. Our specific STM implementation is shown in Figure~\ref{fig:stmpic}. 

To ensure that the STM reaches thermal equilibrium quickly over a wide range of temperatures, the microscope body was made of metal.  Specifically, we used titanium due to its good thermal expansion match with Macor and PZT, the other two materials prominently used in the construction of the STM head. Additionally, titanium is non-magnetic, allowing for future use of the microscope in a magnetic field environment.

To provide wider-area capability than the microscope's tube scanner, an XY coarse translation stage, utilizing a similar design\cite{hug} was built. While the stick-slip nature of the motion prevents such a positioner from being used for scanning, its 3-millimeter range adds versatility by allowing one to locate and study lithographically defined structures or interesting areas in a nominally inhomogeneous sample. In the former case, however, the sample surface will need to be conducting at all points and alignment marks have to be defined to help find the patterns.  

The waveforms required to drive the six shear piezo stacks of the XY and Z coarse motion stages are generated by a dedicated mixed digital/analog circuit, amplified up to $\pm$200V by Apex Microtechnology PA85 high voltage op-amps.  Apex PA88's were utilized in the X and Y high voltage amplifiers used for scanning, with a PA85 chosen to drive the Z piezo for its higher speed. The $\pm$200V maximum output of the piezo drivers gives the STM a 15~$\mu$m scan range at room temperature, dropping by a factor of five at 4K.

\subsection{\label{cryo}Cryogenics}
\begin{figure}[h]
\includegraphics*[width=.5 \textwidth]{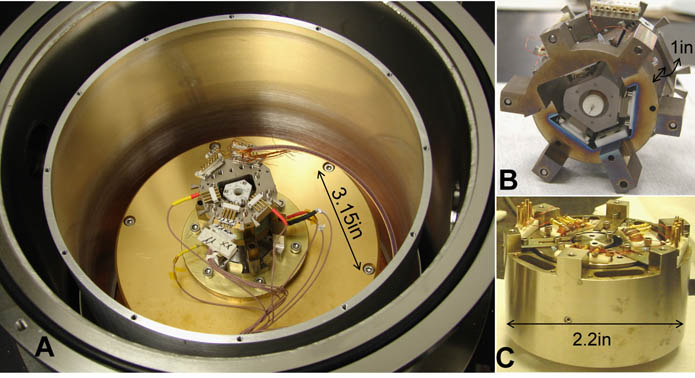}
\caption{Cryostat and STM photos. a) STM mounted inside the cryostat; Z motor and its connector plate are seen from the top. b) Z motor seen from from the tip looking up. c) XY motor.}
\label{fig:cryopic2}
\end{figure}

To achieve short turnaround times during cryogenic operation, the homemade STM described above was mounted on a commercial flow cryostat, the Desert Cryogenics TTP4 Probe Station\cite{desert}.  This cryogenic system consists of a single 9.7 inch I.D. vacuum chamber, with a temperature controlled 6.85 inch diameter sample stage, cooled by liquid helium or liquid nitrogen flowing through a heat exchanger. This stage is surrounded by an 8.4 inch I.D. radiation shield (seen in Figure~\ref{fig:cryopic2}), with its own thermometer and heater. The relatively large TTP4 system has sufficient space to permit the use of a spring-based vibration isolation system for the STM, but as will be shown later, so far the flowing helium has not caused any degradation of performance due to vibrations. As seen in the figure, the STM is simply bolted down to the sample stage, providing good thermal contact. Coax wiring was used for sample and STM tube scanner connections, with twisted pairs for coarse motion wiring, and a triax for the tunnel current wire.  All leads and associated heat sinking were included in the construction of the system by the manufacturer. The cryostat is mounted on a stand with compact pneumatic isolators\cite{newport}  for overall vibration isolation, although these were not floated for the experiments described here. Prior to any cooldown, the system is pumped down with a turbopump to the 10$^{-5}$ Torr range. The pump is then disconnected to avoid vibrations.

\section{\label{nonhardware}Electronics and control}
\subsection{\label{scheme}STP Measurement Scheme}
\begin{figure}[h]
\includegraphics*[width=.5 \textwidth]{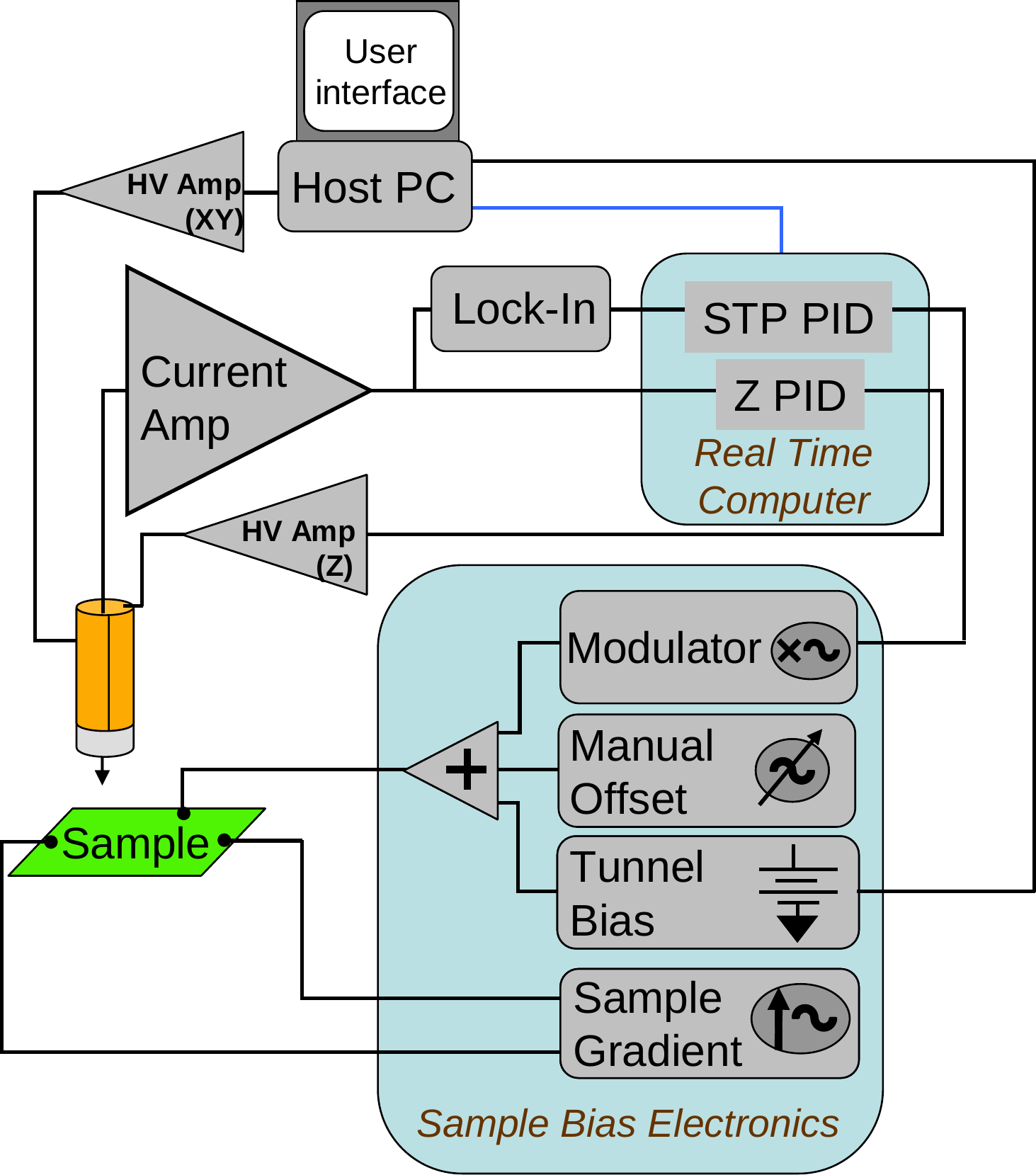}
\caption{STP measurement system diagram. STM and STP feedback loops are implemented digitally on a dedicated processor. STP feedback loop is demodulated by using a lock-in amplifier. The gradient across the sample is established by a floating AC current source, while the current-nulling feedback signal, a manual AC offset (used to reduce the dynamic range required of the STP feedback loop), and the DC bias needed for STM feedback are added at a separate contact to set the voltage of the sample with respect to ground. With the local sample voltage being effectively measured by the STM tip, the configuration is equivalent to the common 4-point geometry, eliminating effects of contact resistance. }
\label{fig:schemepic}
\end{figure}

The STP feedback loop used in our instrument was shown schematically in Figure~\ref{fig:stpfig}. A more detailed electronic schematic is shown in Figure~\ref{fig:schemepic}.

The overall measurement scheme relies on a mix of low noise analog electronics for sample biasing and signal sensing, and digital computer-based control to effectively combine the STM and STP functionalities.  PID control for both the STM's Z-feedback and the nulling STP feedback is provided by a dedicated real-time processor (National Instruments PXI 8175RT embedded controller), running software implemented using LabView RT. All-digital PID control eases implementation of such critical features as sample-and-hold, and in the future will enable optimization of the speed of the STP feedback loop by using more advanced algorithms beyond simple PID. 
The Z and STP control loops receive their process variables (tunnel current and lock-in output, respectively) from the analog inputs, calculate the PID response for the Z piezo and STP, and then outputs the values using the analog output channels. The feedback controller, which can maintain stable control at loop rates of up to 10KHz, continually broadcasts data via an Ethernet connection to the ``host'' computer, which is a standard PC desktop. The host computer provides user interface and executes the complete STP data acquisition algorithm (also implemented in LabView). To do this, the host can send PID gains and feedback state parameters to the real-time computer, and can set STM tunnel bias and XY scan position by using analog outputs of an on-board NI PCI6733 card.

To eliminate effects of lead and contact resistance variations, we bias the sample in a manner designed to achieve a configuration functionally equivalent to a common 4-point measurement geometry. The gradient across the film is provided by a battery powered floating AC current source while a third contact allows the sample potential to be shifted with respect to ground by a combination of the three signals - the DC tunnel bias, the manual AC offset, and the STP zeroing feedback signal. The fourth ``contact'' is the STM tunnel junction. With the tip current zeroed by STP feedback, no current can flow through the voltage contact, causing the voltage under the tip to be independent of any contact resistance.  Tip current is detected using a Keithley 427 current amplifier (set to the 10$^9$ gain), which is connected, in turn, to an SRS850 lock-in amplifier, with the sample gradient signal as its reference.  The amplitude of the tunnel current at the current bias frequency and phase is used as the process variable for the STP PID feedback loop, the set point of which is zero volts. 

The outputs of the current amplifier and the lock-in are continuously sampled by the analog inputs of a National Instruments PXI 6030E DAQ board installed on the real-time computer's chassis, and are passed on to the topography and STP PID subroutines, respectively. The topography PID output, recorded as Z information, is connected to the Z-piezo of the STM through the high-voltage amplifier. The output of the STP PID is multiplied (using an analog multiplier) by a fixed amplitude reference AC signal (at the same frequency and phase as the bias current), and added to the sample at the voltage terminal to complete the AC potentiometric feedback loop. STP information is then directly given by the output of the STP PID, once the control loop has stabilized. The demodulated feedback scheme greatly improves stability, by permitting the STP feedback loop to operate at a low frequency while controlling higher frequency signals.  A simpler, high bandwidth feedback scheme was also implemented, in which PID uses the tunnel current itself as a process variable, and a lock-in placed outside of the loop is used to obtain STP data. However, the resulting high gain-bandwidth product loop proved difficult to stabilize.

The two other signals that are added to the sample at the voltage contact are not a part of the STP feedback loop. The DC tunnel bias provides the tunnel current for normal STM operation; it is set to zero during acquisition of STP data to reduce vibration-induced noise. Also, a manual offset can be used to bring the AC voltage under the tip close to zero by hand, using analog electronics only. This is desirable to reduce the dynamic range required of the STP feedback circuit. The offsetting voltage is generated simply by using a potentiometer to attenuate the inverted reference AC signal. Manual zeroing is performed only once, at the beginning of a scan, the remainder of which is performed entirely by the host computer.

\subsection{\label{protocol}Measurement protocol}

\begin{figure}[h]
\includegraphics*[width=.48 \textwidth]{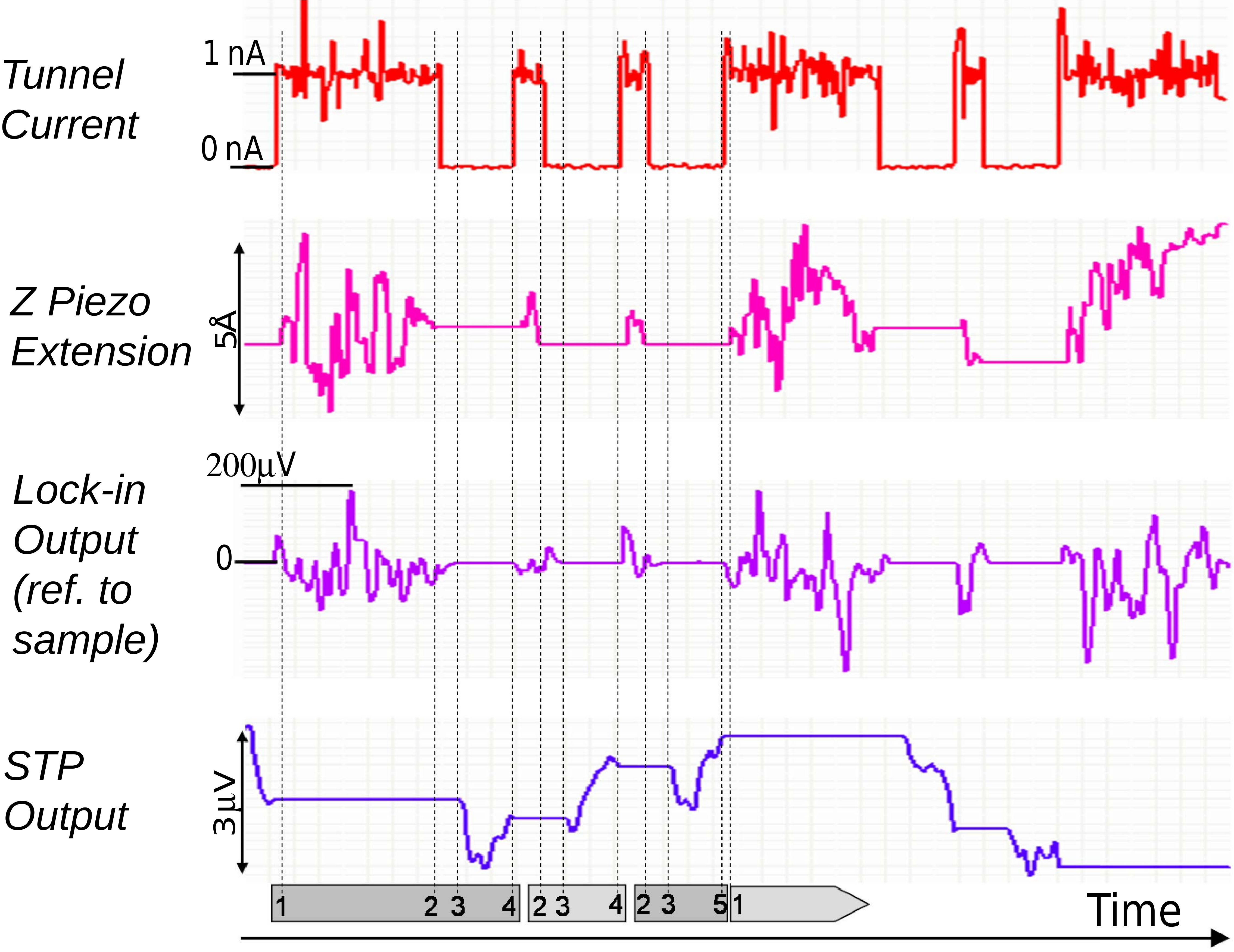}
\caption{STP measurement protocol. Relevant signals shown as a function of time to demonstrate sequence of events. The detailed sequence indicated on the bottom is described in detail in the text; briefly, the events are: 1 - move XY, 2 - open Z loop, 3 - close STP loop, 4 - open STP loop, close Z loop, test convergence and reject attempt, 5 - accept point. Signal range shown is representative, and depends highly on scan parameters. }
\label{fig:protocolpic}
\end{figure}

Images are taken by scanning the tip in a raster pattern, obtaining topographic and potentiometric information at each pixel. The sequence of events and the associated signals involved in acquiring a single pixel of Z and STP data are demonstrated in Figure~\ref{fig:protocolpic}, which shows actual time traces of tunnel current, Z-piezo extension, lock-in output, and STP PID output. At time index 1 (see bottom of figure), the microscope initiates a lateral step: the Z feedback loop is closed, keeping the tunnel current at its set point, while the STP feedback loop is open, holding its previous value, to avoid subjecting the loop to the obviously noisy process variable. At the completion of the step (index 2), the Z-feedback loop is opened, holding Z constant, and the DC tunnel bias is set to zero, bringing the DC tunnel current to zero as well. The lock-in noise also drops significantly, as the noise caused by vibrations at the gradient frequency is eliminated once all tip-sample voltages are zeroed. After a short wait for the transients to decay, the STP feedback loop is closed (index 3).

Ideally, the system would be kept in this state (STP loop closed, Z loop holding) until the lock-in output in the potentiometry loop converges to zero; however, this step is typically limited to a couple of seconds to avoid piezo-creep-induced tip crashes while the Z feedback is off. Since the loop time constant is typically set to one second, the feedback may (and often does) fail to converge in a single attempt. The STP feedback loop is then opened (index 4), the tunnel bias is restored to a finite value, the Z feedback loop is closed, and the junction is allowed to stabilize. 

STP loop convergence is then evaluated by checking whether both the process variable (lock-in output) error and control parameter (STP output) variation over the loop-closed interval lie within a specified range. In the figure shown, the parameters were set to always reject the first attemp at a new location, so steps 2-4 are repeated twice more. On second attempt, the loop is deemed to have not yet converged (change in STP from beginning to end of the attempt exceeding threshold). At index 5 (third attrempt), the STP/Z data is accepted, and the tip is moved to a new location to start on the next pixel. In practice, convergence criteria are picked so that a point is typically accepted in fewer than five attempts. 
If the loop does not converge after a pre-set number of attempts, the STP point is skipped and only topographical data is recorded. 

STP points may be skipped for another reason - if a sufficiently stable tunnel junction cannot be maintained during STP acquisition at any given location. This is done to prevent the STP loop from deviating far from the true STP value, since the higher gain and wider bandwidth associated with a low junction resistance can destabilize the system, especially if the tip-sample voltage is not yet zeroed. Although the junction resistance cannot be monitored during STP acquisition since the DC tunnel bias is off, it can be checked immediately after. When the DC tunnel bias is restored to its finite value after opening the STP feedback loop, but before the Z feedback loop is re-established (just before step 4), the junction is evaluated by comparing the tunnel current to the set point. If the junction resistance has drifted too far (typically by more than factor of two) during the Z-feedback hold, the attempt is ``forgotten'' by restoring the STP output to its pre-attempt value. If a stable junction cannot be attained after a fixed number of attempts, the pixel is skipped as in the failed convergence case.

With a measurement time constant of one second, five attempts per point, and some overhead time spent stabilizing the junction and switching voltages, a 100x100 pixel STP image takes about 24 hours. This places significant demands on long-term thermo-mechanical stability of the STM head and thermo-electrical stability of STP electronics, and increases motivation for optimizing the feedback system for faster convergence time.  In the future, the considerable flexibility of the real-time computer feedback system along with the detailed characterization of the transfer characteristics of the other components presented below could and should be used to optimize the system globally and reduce the overall scan time.

\subsection{\label{electronics}Sample biasing electronics}

Let us now turn to the sample biasing electronics.  Besides the need for low thermal drifts and low-noise performance, another requirement is good phase alignment between the signals that cancel each other during the STP measurement.  These include the sample bias current, the manual offset and the STP feedback signal. While lock-in detection allows only the in-phase signals to be used for potentiometry,  it is still important to minimize any quadrature signals that might interfere with normal STM operation. Additionally, poor phase alignment of the large canceling signals would make the system more susceptible to noise caused by possible phase fluctuations introduced by cable microphonics. 	

\begin{figure}[h]
\includegraphics*[width=.5 \textwidth]{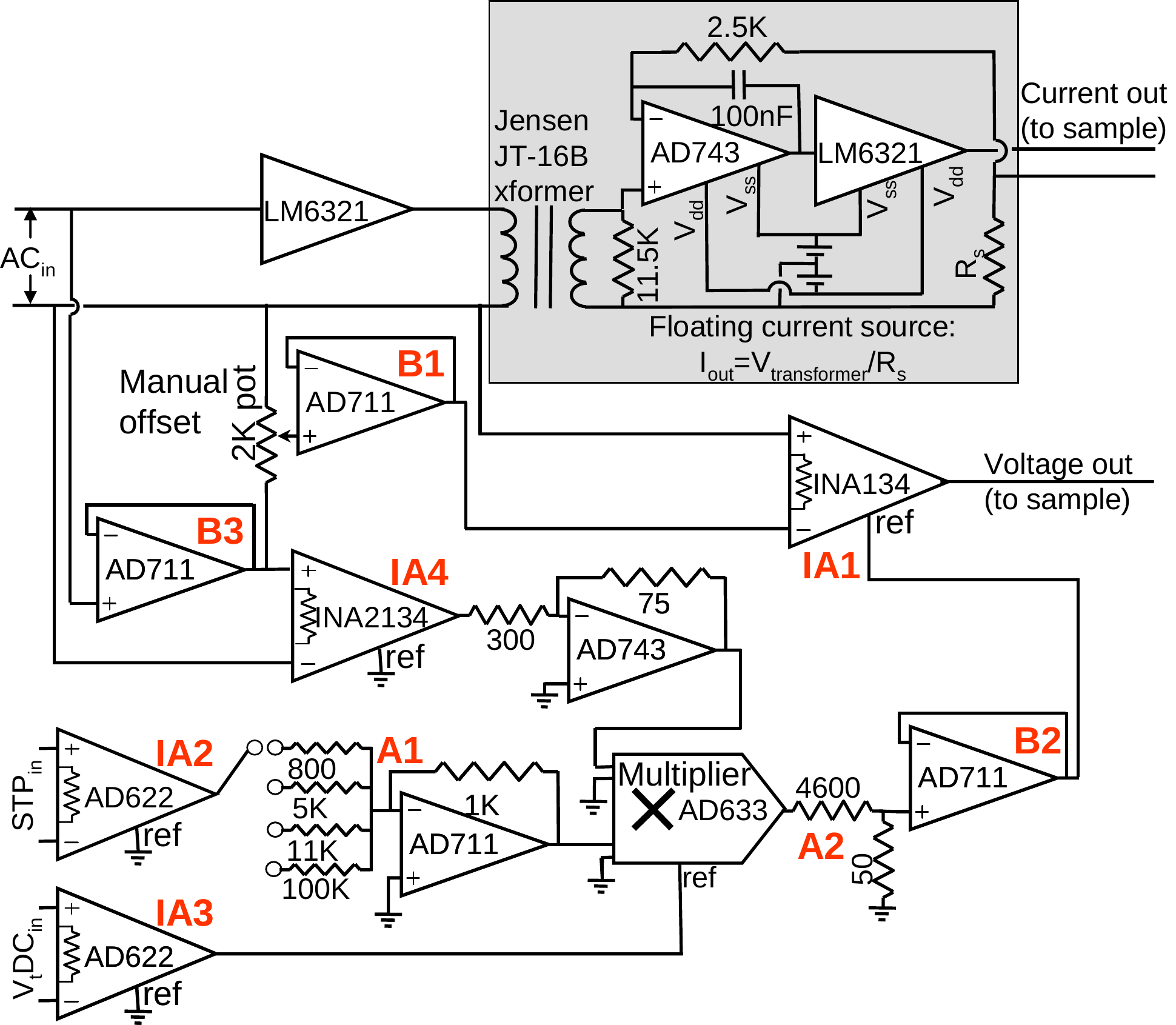}
\caption{Sample biasing electronics diagram. A single, constant AC reference AC$_{in}$~drives all AC signals connected to the sample: sample current bias (generated by the floating voltage-controlled current source),  tip-current-nulling feedback voltage (generated by the STP feedback signal multiplying the reference), and the manual offset (generated by voltage-dividing the reference signal).}
\label{fig:circuitpic}
\end{figure}

The diagram of the analog circuit used to bias the sample is shown in Figure~\ref{fig:circuitpic}. As noted above the AC bias across the sample is established by a completely floating current source, while a voltage with respect to a defined ground is set at a third contact. The circuit is thus composed of two separate parts: the current source, powered by two rechargeable 12V batteries, and the voltage setting section, powered by a standard power supply connected to a proper ground. The circuit chassis, cryostat and the virtual ground of the current amplifier all share the same ground, making it the single common reference for the measurement. 

The current is supplied by a single op-amp voltage controlled current source for a floating load \cite{apex}, with an LM6321 driver, powered by the same battery, used to boost (to 200~mA) the maximum current output to enable biasing of low-resistance samples. Since a floating current source is susceptible to oscillations, a feedback RC roll-off network is needed to stabilize it. A transformer is used to couple a 237~Hz signal from the lock-in reference output to the floating part of the circuit. A second LM6321, in the non-floating section of the circuit, is used to drive the transformer primary. A current setting resistor is picked to maximize sample gradient, limited by the driver compliance or sample heating. Since the current is determined by this single resistor, very stable Vishay S102C series foil resistors with a low temperature coefficient of 2ppm/$^{\circ}$C are used.

In the non-floating part of the circuit, an instrumentation amplifier (IA1) adds the voltage setting signals together. Other instrumentation amplifiers (IA2-IA4) are used at the input stages for the STP PID and DC tunnel bias signals from the DAQs, as well as the fixed AC signal from the lock-in. The amplifiers isolate the circuit from the grounds of these instruments. Selectable attenuator (A1) can be used to lower the effects of noise of the STP DAQ output, at the expense of reducing the dynamic range of the STP feedback. The relatively high 0.8~$\mu$V/$\sqrt{Hz}$ output noise of the AD633 multiplier modulating the STP signal is reduced by the attenuator A2. Manual offset generated from the reference AC signal with a single Clarostat 73JA pot used as a voltage divider.

Since all resistors in the non-floating part of the circuit are used in a voltage divider configuration, this configuration cancels any temperature dependence as long as the temperature coefficients are the same. Simple metal film resistors are sufficient. Followers (B1 and B2) are used as buffers between the dividers and instrumentation amplifiers to prevent the loading of the dividers by the input resistors of the instrumentation amplifiers (which would eliminate resistor temperature coefficient matching and lead to high thermal drifts).

\section{\label{performance}Component performance}

\subsection{\label{stmperf}STM}

Performance of any scanning tunneling microscopy system can in general be characterized by the vibration noise of the tip-sample separation. Although, as described above, scanning potentiometry is generally immune to tunnel resistance variations caused by vibrations, the same is not true for basic STM imaging. Additionally, as will be discussed later, some STP applications do require good vibration performance as well. For the system being described here, an important issue is whether the flow cryostat, with its expected vibrations, is suitable for such a sensitive instrument.
	  
\begin{figure}[h]
\includegraphics*[width=.45 \textwidth]{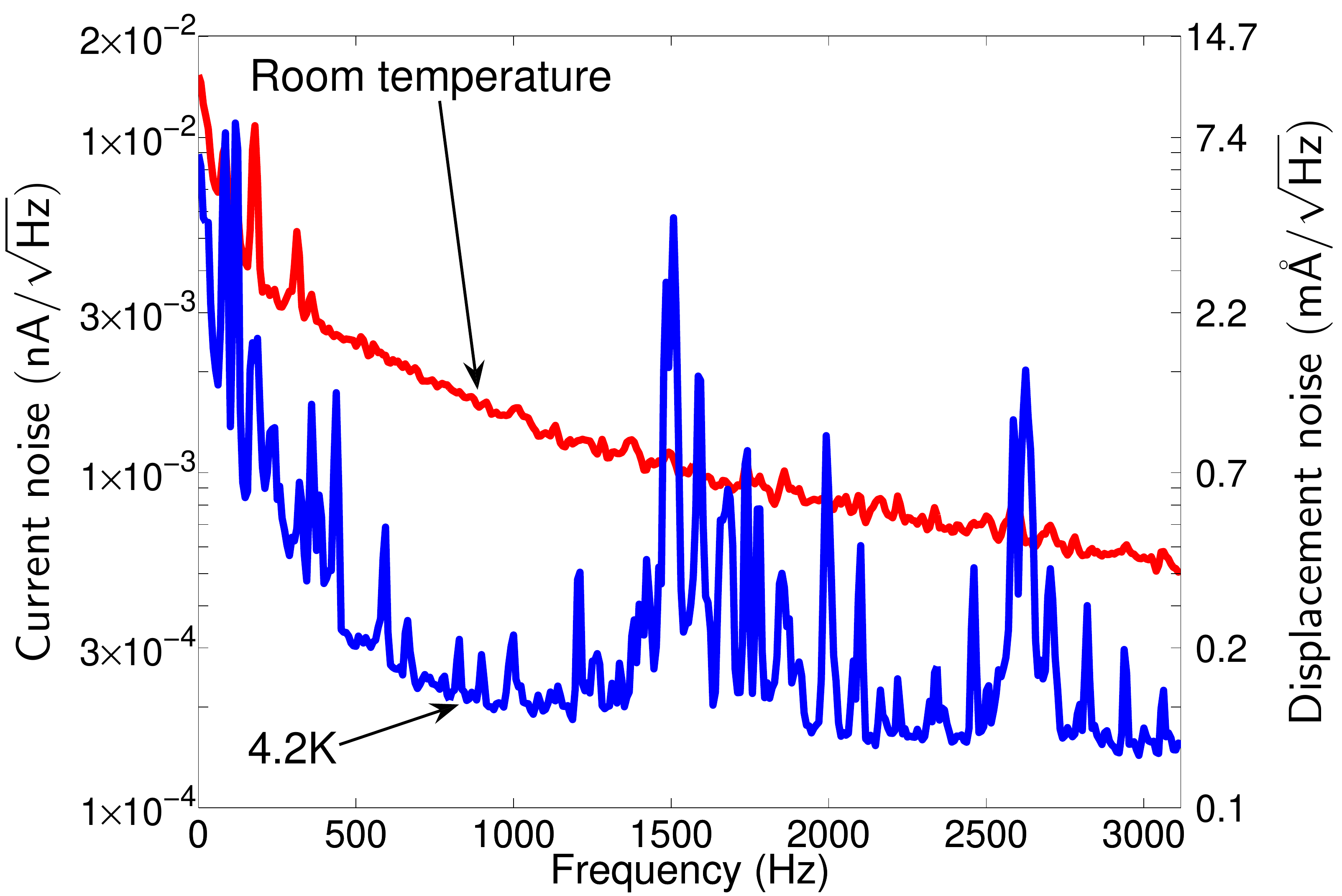}
\caption{STM current and associated displacement noise for 4~nA setpoint, 400~mV tunnel bias. Current noise converted to approximate displacement noise by using I=I$_{0}$e$^{-0.34z}$ (I$_{0}$=4~nA, z in \AA), obtained from separate dI/dz measurement performed at room temperature.}
\label{fig:vibrpic}
\end{figure}

The amplitude of the tunnel current noise, relative to the current set point, can be used to extract the spectral density of the vibrations. For this measurement, scanning is disabled and STM PID gains are turned down. The resulting low feedback bandwidth is then able only to prevent tip crashes caused by slow processes (thermal drift and piezo creep), but is not fast enough to respond to any other sources of current noise. Figure~\ref{fig:vibrpic} shows the spectral density of the current noise obtained by processing the STM current signal with an FFT spectrum analyzer. Spectra are shown for room temperature in air and for 4.2~K in vacuum. The current noise is seen to be quite low compared to the 4~nA current set point. Equivalent displacement noise, determined from the separately measured dI/dz characteristic of the junction, is shown on the right Y-axis.   

As seen in the figure, tip-sample displacement noise is on the milliangstrom/$\sqrt{Hz}$ level or lower.  In ambient conditions, the spectral density shows 1/f-like behavior with some 60Hz harmonics peaks. At 4.2~K, the 1/f envelope drops significantly, revealing lower-level 60Hz harmonics. New, broad peaks of larger magnitude than room-temperature levels also appear near 1500 and 2500Hz. We interpret the 1/f noise as thermally activated motion of impurity atoms on the sample or the tip. At 4.2~K this motion is frozen out, significantly lowering this source of noise. The 1500/2500Hz peaks are the mechanical resonances of the STM head itself, driven either by the helium flow or by external vibrations transmitted along the long transfer tube connected to the cryostat. However, as will be shown below, such small external vibrations do not affect imaging performance of the microscope. Additionally, due to their high frequency, these vibrations can be eliminated with the most rudimentary spring-based isolation system if that were to become necessary for a more vibration sensitive applications, such as scanning tunneling spectroscopy.

\subsection{\label{capacitance}Tip-sample capacitance}

To correctly probe the electrochemical potential under the tip under AC bias conditions, the tunneling resistance must be the dominant conduction path between the tip and the sample. With the STM just out of the tunneling range, the tip-sample transfer characteristic is consistent with a pure 0.08~pF capacitance, combined with the multipole roll-off of the current amplifier above 500~Hz. At our measurement frequency of 237 Hz, below the roll off, the capacitive reactance of 10$^{10}$~$\Omega$ is shorted out even by the highest tunnel resistances used (100~M$\Omega$), with much stronger rejection due to the 90$^{\circ}$ phase shift of the capacitive signal.  

Moving the tip from just outside the tunnel range (a few angstrom away) to a distance of 700~nm causes a drop in the capacitive coupling of less than 1\%. This indicates that only a small part of the measured capacitance comes from the immediate vicinity of the tip-sample junction. The low spatial sensitivity of the already small capacitive signal, combined with sub-Angstrom vibrations of the tip, eliminates any possibility of vibration induced capacitive noise, which would actually be back in phase with the voltages on the sample. Therefore, the STP system cannot inadvertently function as a Kelvin probe driven by tip vibration or as any other probe of electrostatic potentials. 

\subsection{\label{pid}Digital PID control}

\begin{figure}[h]
\includegraphics*[width=.48 \textwidth]{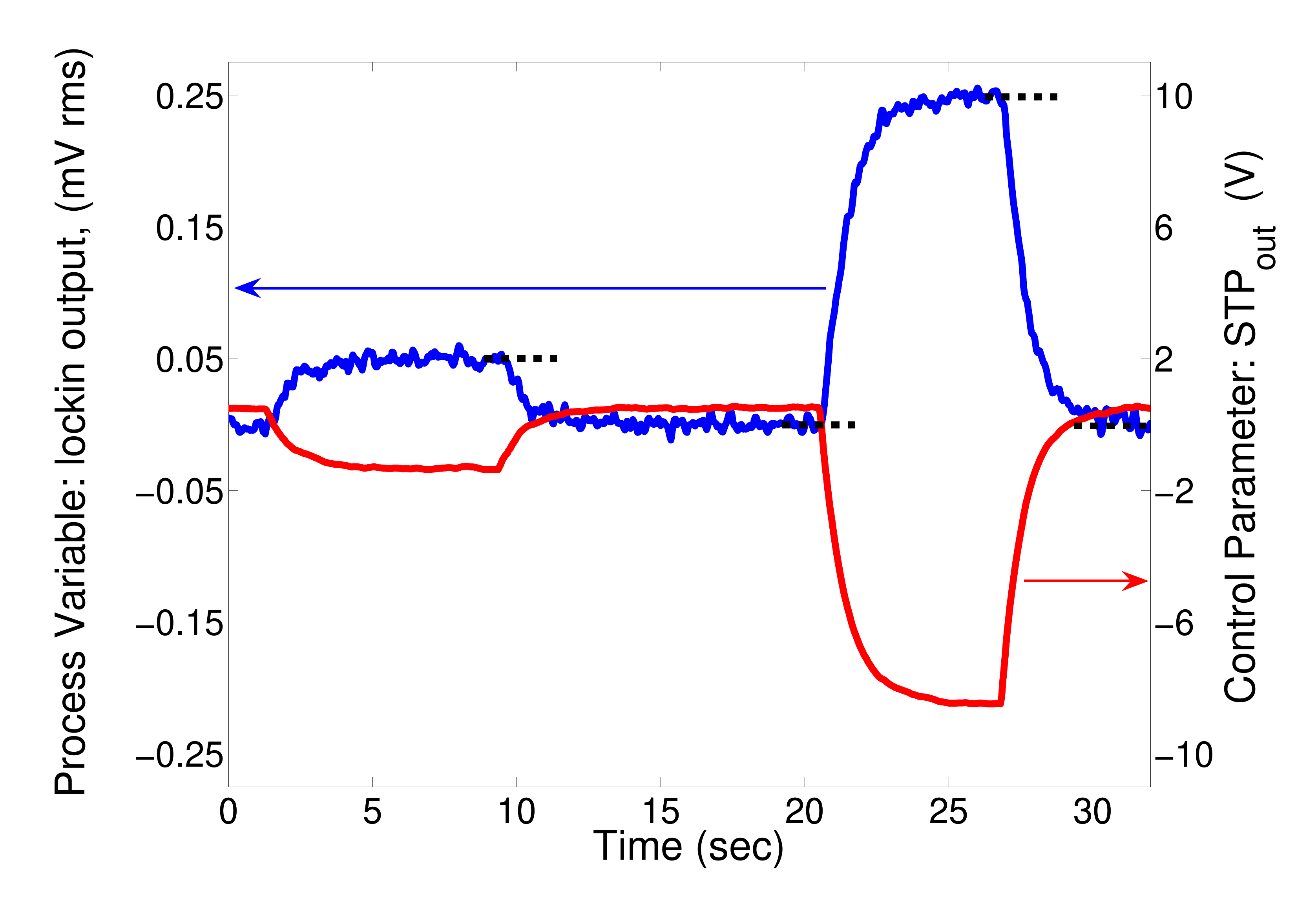}
\caption{Step function response of the STP feedback loop, measured with junction and sample replaced by resistors. Feedback system is convergent, with observed time constant equal to calculated value (1 sec). Dashed lines represent setpoints of the feedback loop.   }
\label{fig:pidpic}
\end{figure}

	Closed loop characteristics of the STP electronics were tested to verify the functionality of the demodulated feedback scheme. The STM head and the sample were replaced by resistors -- a large carbon resistor to simulate the tunnel junction, and smaller resistors for the ``sample''. In these measurements, only the integral component of the PID was used for simplicity; however, full functionality should be used later to improve overall loop convergence time. The step-function response of the STP PID loop, measured by changing the setpoint (represented by dashed lines) of the feedback loop, shows (Figure~\ref{fig:pidpic}) convergent behavior, with observed time constants consistent with calculated values. Stability was achieved by maintaining a single dominant pole in the loop - keeping the lock-in time constant much shorter than that of the integrator; the time constant difference can be seen in signal-to-noise difference between the lock-in and STP traces. One peculiarity of the STP loop involving an actual tunnel junction is the increase of process variable noise with process variable error (i.e. when there is a finite voltage across the junction). It is possible that, for such a system, loop convergence time can be improved by control schemes more sophisticated than an integrator or full PID.
	
\section{\label{initialmeas}Initial measurements}

\begin{figure*}[tb]
\includegraphics*[width=1 \textwidth]{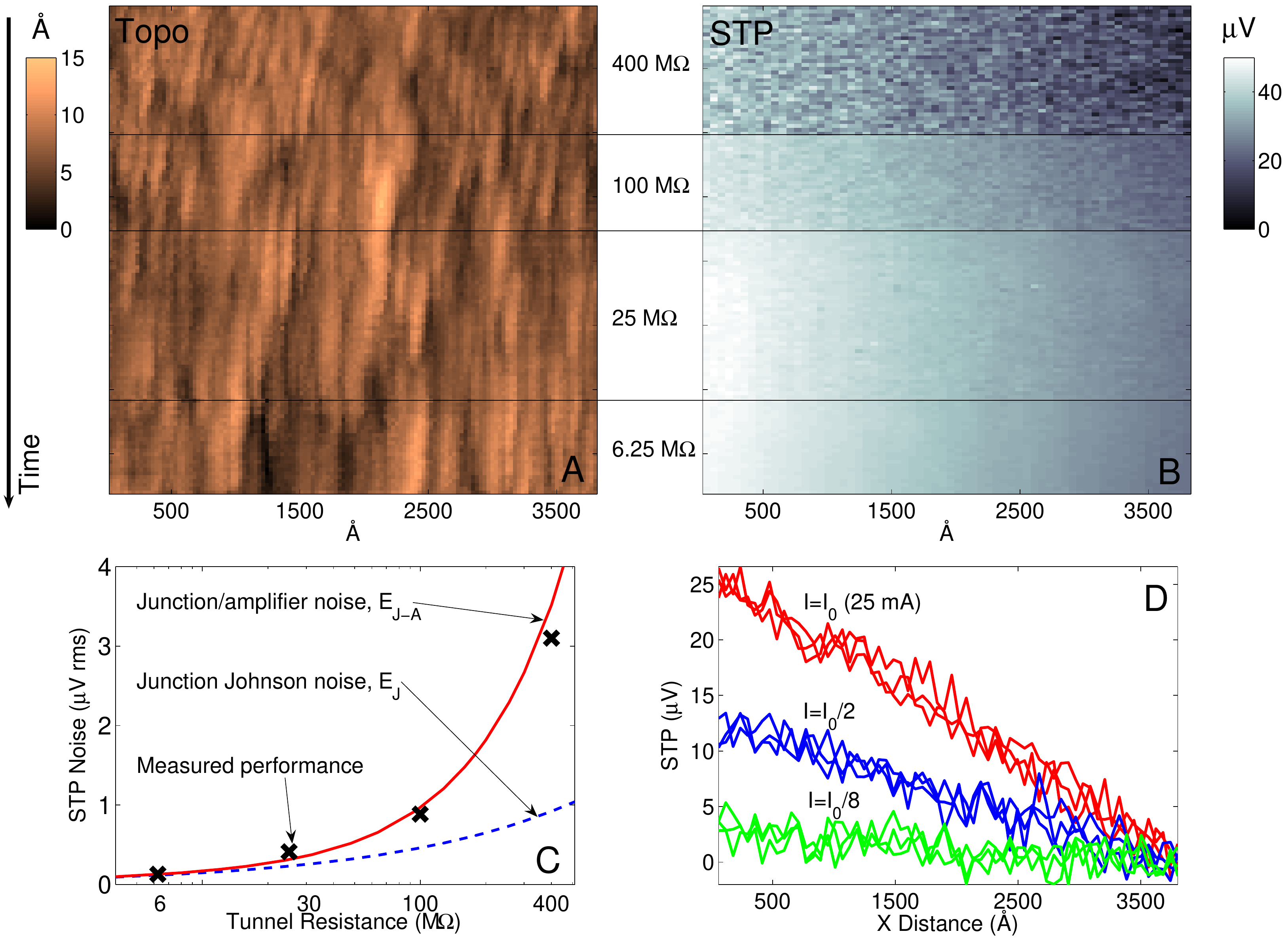}

\caption{ Demonstration of the functionality of STP. a) Successive topographic traces of a single scan line on the sample as a function of time (top to bottom) for various values of junction resistance.  The value of the junction resistance is shown in the central column between a) and b).  b) Corresponding potentiometric scanlines to the topographic ones in a). The data show reduction of potentiometry noise with decreasing junction resistance with no degradation to the quality of the topographic
image. c) Measured and predicted STP noise vs. junction resistance.  The data show that the fundamental noise limit is reached as the junction resistance is decreased. d)  STP scans for various bias currents.  Note that the observed expected scaling demonstrates that the observed signals are indeed the electrochemical potential. This and all following figures, with the exception of Figure~\ref{fig:cooldownpic}, show data taken at room temperature.
}
\label{fig:basicpic}
\end{figure*}

\subsection{\label{basic}Basic functionality}
A thin, 20~\AA~gold film, deposited on top of a 10~\AA~Ni wetting layer, was used to demonstrate a complete STP measurement, the results of which are shown in Figure~\ref{fig:basicpic}. Biasing the sample with a 25 mA rms current sets up a 1.4~V rms voltage measured across the 5 mm sample and its wiring, with about 2/3 of resulting voltage drop occurring at the wirebonded contacts (estimated to be 15-20~$\Omega$/contact from separate measurements). The long length scale average electric field is therefore on the order of 1~V/cm. The observed gradient, shown in Figure~\ref{fig:basicpic}d, is 0.7~V/cm, consistent with the known applied field; this agreemement, as well as the scaling of the measured gradient with the applied current (also shown in the figure), confirm that the system is in fact measuring a transport-related electrochemical potential. 

To verify that the STP measurement reaches the fundamental Johnson noise limit, several tunnel junction resistance values were used while keeping the STP loop time constant at 2 sec (corresponding to 1/8~Hz equivalent noise bandwidth [ENBW]). Figures~\ref{fig:basicpic}~a,b show topographic and potential ``images'' taken by scanning the same scanline repeatedly, although the start of temperature control at the top of the image causes spatial drifts to settle only toward the end. Drifts are addressed in Section~\ref{longterm}. Here and later in the paper we obtain the STP voltage noise by calculating an rms average of pixel-to-pixel differences of all pairs of successive scanlines rather than by keeping the tip stationary, in order to verify the simultaneous imaging capability. 

As can be seen from Figure~\ref{fig:basicpic}c, the drop in the STP noise as the tunnel resistance is lowered is reasonably well described by a simple model of the thermal noise (E$_{J})$ of the junction resistance combined with the 24~fA/$\sqrt{Hz}$ current amplifier noise (E$_{A})$, with the total predicted noise (E$_{J-A})$ given by 
\begin{align*}
				&E_{J-A}=\sqrt{E_{J}^{2} + E_{A}^{2}},\\
        &E_{J}=\sqrt{4k_{b}\times 300K\times R_{t}\times (ENBW)}, \\
        &E_{A}=(24fA/\sqrt{Hz})\times R_{t}\times \sqrt{ENBW}.      
        \label{Ej}
\end{align*}
Referenced back to the sample, the amplifier noise is multiplied by the tunnel resistance. Therefore, at lower junction resistances, the measurement is limited by the junction Johnson noise. At cryogenic temperatures, however, with the junction thermal noise reduced, the amplifier noise can become the noise-limiting parameter of the system, as will be described in Section~\ref{cryoop}.

The 130~nV/$\sqrt{Hz}$ noise measured for a tip-sample resistance of 6~M$\Omega$ is the lowest published to date. As far as topography is concerned, no image degradation is seen to occur as the junction resistance is lowered by two orders of magnitude, although other materials will have different minimum junction resistance thresholds for acceptable STM operation.

\begin{figure}[h]
\includegraphics*[width=.5 \textwidth]{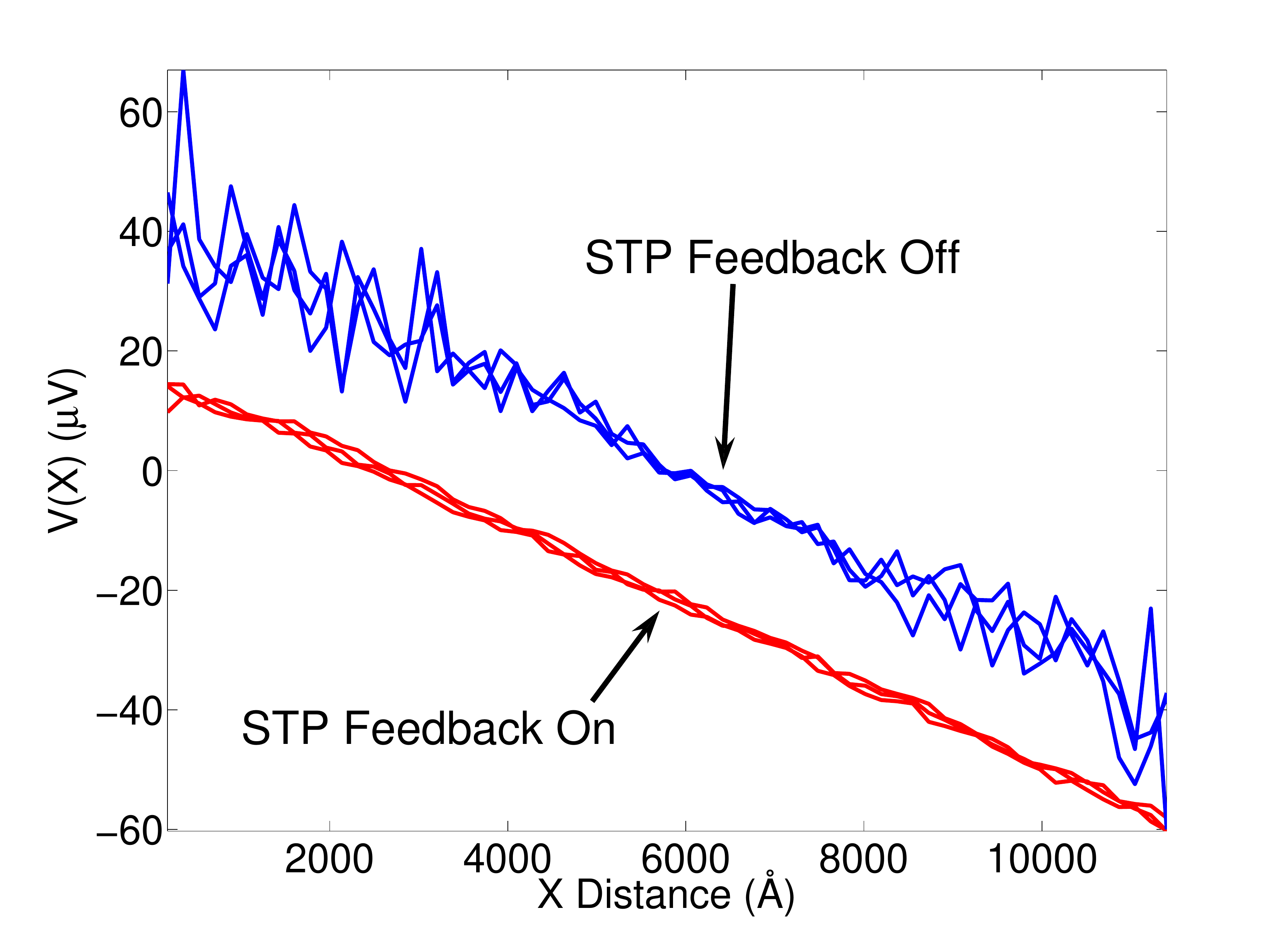}
\caption{The importance of potentiometric feedback: without continuous zeroing, noise increases linearly as tip-sample voltage increases at large distances from initial zeroing point.}
\label{fig:stpfeedbackpic}
\end{figure}

As alluded to above, the only system that reached Johnson-noise-limited STP performance\cite{pelzlownoise} (also by using AC techniques), did not involve using active feedback to continuously null the tip-sample voltage. At such low noise levels, however, the effect of finite voltages that develop at larger scan sizes become important. Figure~\ref{fig:stpfeedbackpic} illustrates this problem and its solution by comparing potential line scans taken on a thin gold film using STP feedback with scans for which the feedback loop was kept open after a single zeroing of the current. Potentiometry data taken with no feedback, obtained from the lock-in output by using the nominal tunnel resistance setpoint and current amplifier gain, show the correct gradient, with noise increasing rapidly as the tip-sample voltage increases beyond 10~$\mu$V at larger distances from the initial zeroing point. The extra noise is caused by tip vibrations and would be a limiting factor on a useful scan range even for homogeneous transport studies.  It would also hamper any applications on materials dominated by inhomogeneities, where local fields can be much higher and potentials change rapidly. While improving the vibration performance of the STM can reduce noise sufficiently for some specific measurements, STP feedback resolves the problem completely, making it the ideal solution for a versatile instrument.

\subsection{\label{finitedc}Finite DC bias}

Although keeping the tip-sample voltage at zero is instrumental for achieving Johnnson noise-limited performance, for some non-linear systems maintaining a finite DC bias during the STP measurement is required. One example is the study of superconductors, where biasing above the superconducting gap may be needed to keep the tunnel junction resistance reasonably low. Measurements performed to test STP performance under such conditions show that STP noise increases linearly with tip-sample DC bias once the Johnson noise floor imposed by tunnel junction is exceeded. The slope of the rise of STP noise density is 500~nV/$\sqrt{Hz}$ per 1~millivolt of DC bias (in the case of DC biasing the additional noise is smaller than for a corresponding AC tip-sample voltage, since only vibrations at the measurement frequency can contribute to the noise). Therefore, microvolt noise levels are still possible with our system at DC voltages of several millivolts, above the gap for conventional superconductors. However, to maintain such performance at higher bias voltages, or to achieve lower noise at millivolt bias levels with smaller junction resistances or at lower temperatures, additional vibration isolation will be required. 

\subsection{\label{fourpoint}Effect of contact resistance}

Since the STP system described here can measure local potentials much smaller fractionally than the bias applied across the sample, it is susceptible in principle to even small variations in the lead contact resistances, necessitating the implementation of the four-point measurement scheme described above. The effect of contact resistance variations was tested on a SrRuO$_{3}$ sample with a 200~$\Omega$  resistance simply by putting 300~$\Omega$  potentiometers in series with the three sample leads: the two current contacts and the single voltage lead. The values for the pots, located at the outputs of the STP electronics box, were picked to model the extreme case of contact resistance ranging from zero to above full sample resistance. A full resistance swing of the voltage contact resulted in a 40~$\mu$V change in the voltage under the tip, a small fraction of the 150~mV gradient across the sample. Changing the current leads' contact resistance was found to have no effect. In practice, the biasing scheme implemented to deal with any contact problems avoids the need for depositing dedicated gold contact pads on the samples, with wirebonded contacts sufficient for measurements on different types of films.

\subsection{\label{longterm}Long-term stability}

\begin{figure}[h]
\includegraphics*[angle=90,width=.48 \textwidth]{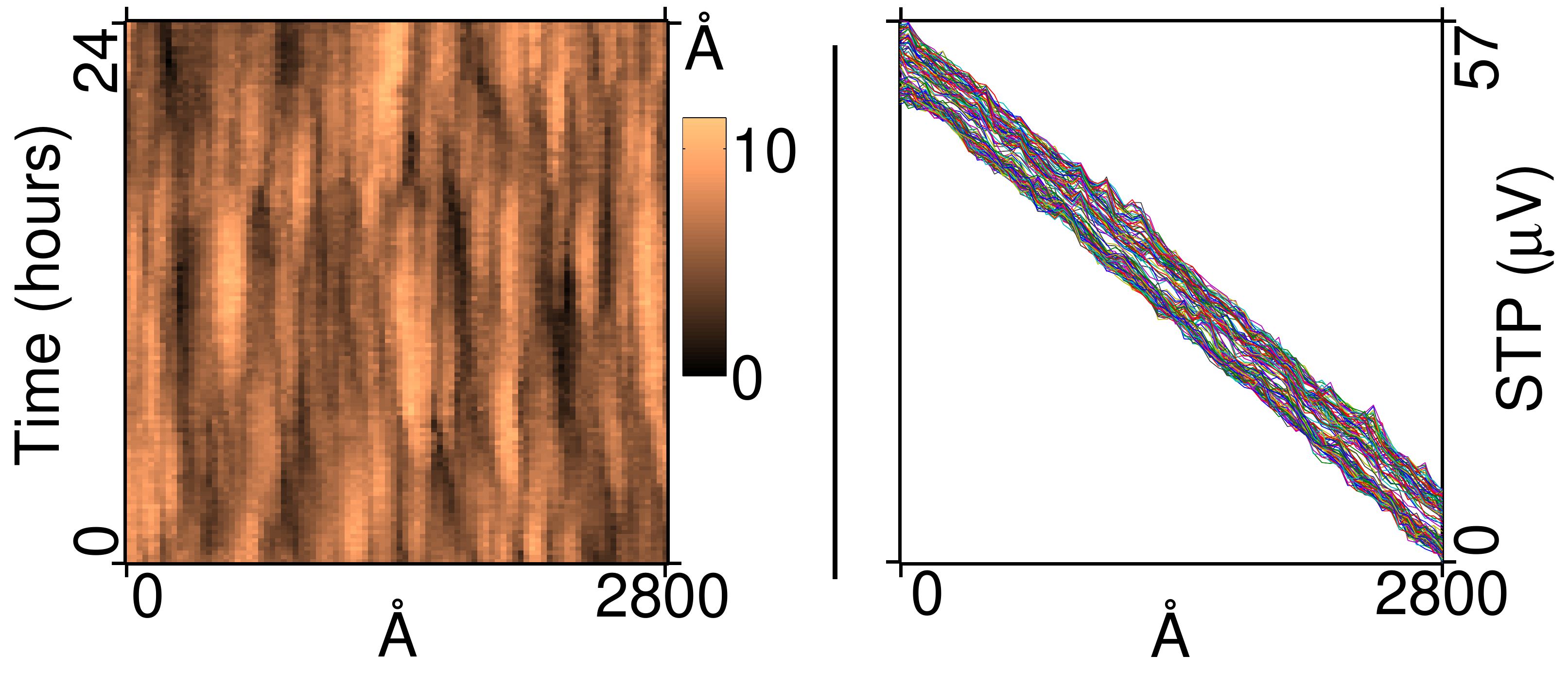}
\caption{Temperature drifts of complete STP system. Left: topography obtained by a 24 hour 2800x0~\AA~STP scan; 5~mK thermal drifts are responsible for lateral drifts of 5-10~\AA/hour. Right: All 128 STP scanlines from the scan; potential drifts do not exceed 8~$\mu$V over 24 hours. 
}
\label{fig:longtermpic}
\end{figure}

Due to the time consuming nature (and complexity) of the STP measurement, the system is particularly vulnerable to temperature drifts. These include spatial drifts of the STM head, electrical drifts in the STP biasing electronics and variations in potentials due to temperature-dependent properties of the sample itself. These drifts are addressed by a combination of temperature control of the STM and the sample, thermally stable design of the microscope head and STP electronics, and orientation of the fast scan axis along the expected maximum field direction.  

Figure~\ref{fig:longtermpic} shows data from a 2800$\times$0~\AA~(same line scanned repeatedly) STP scan taken on a thin gold film with a 2~V/cm applied gradient, taken over 24 hours at room temperature in air. Due to convection, thermal stability is more difficult to achieve under these conditions, but the system is typically pumped out only for lower temperature operation, to avoid the possibility of high voltage arcing of the piezo wiring at intermediate pressures. With ambient temperature varying by about 1$^{\circ}$C peak-to-peak and the radiation shield stage controlled to within 50~mK, the microscope (and the sample) temperatures can be kept constant to within 5~mK. This leads to thermal spatial (lateral) drifts of 5-10~\AA/hour, as determined from the shifts in the topographical image, which would show perfectly vertical features in the absence of any drifts. The drifts are caused by the remaining 5 mK temperature variations and presumably will drop significantly at 4.2~K, although we did not test this directly.

A look at all 128 STP linescans shows the total potential drift is 8~$\mu$V over 24 hours, which may be caused by actual voltage drifts in the electronics, as well as the lateral tip drift along the gradient direction. The scale of the possible voltage drifts is not surprising, since zeroing the potential under the tip involves subtracting the 1~V level gradient signal from the 1~V level manual offset. Therefore, the ppm/$^{\circ}$C magnitude temperature coefficients of components used in the biasing circuits can be expected to lead to drifts of several microvolts in the presence of the observed 1$^{\circ}$C ambient temperature variation. Any further improvement in the long-term stability of the measurement would require temperature control of the STP electronics box itself.

\subsection{\label{creep}Effect of piezo creep at large scan sizes}

\begin{figure}[h]
\includegraphics*[width=.5 \textwidth]{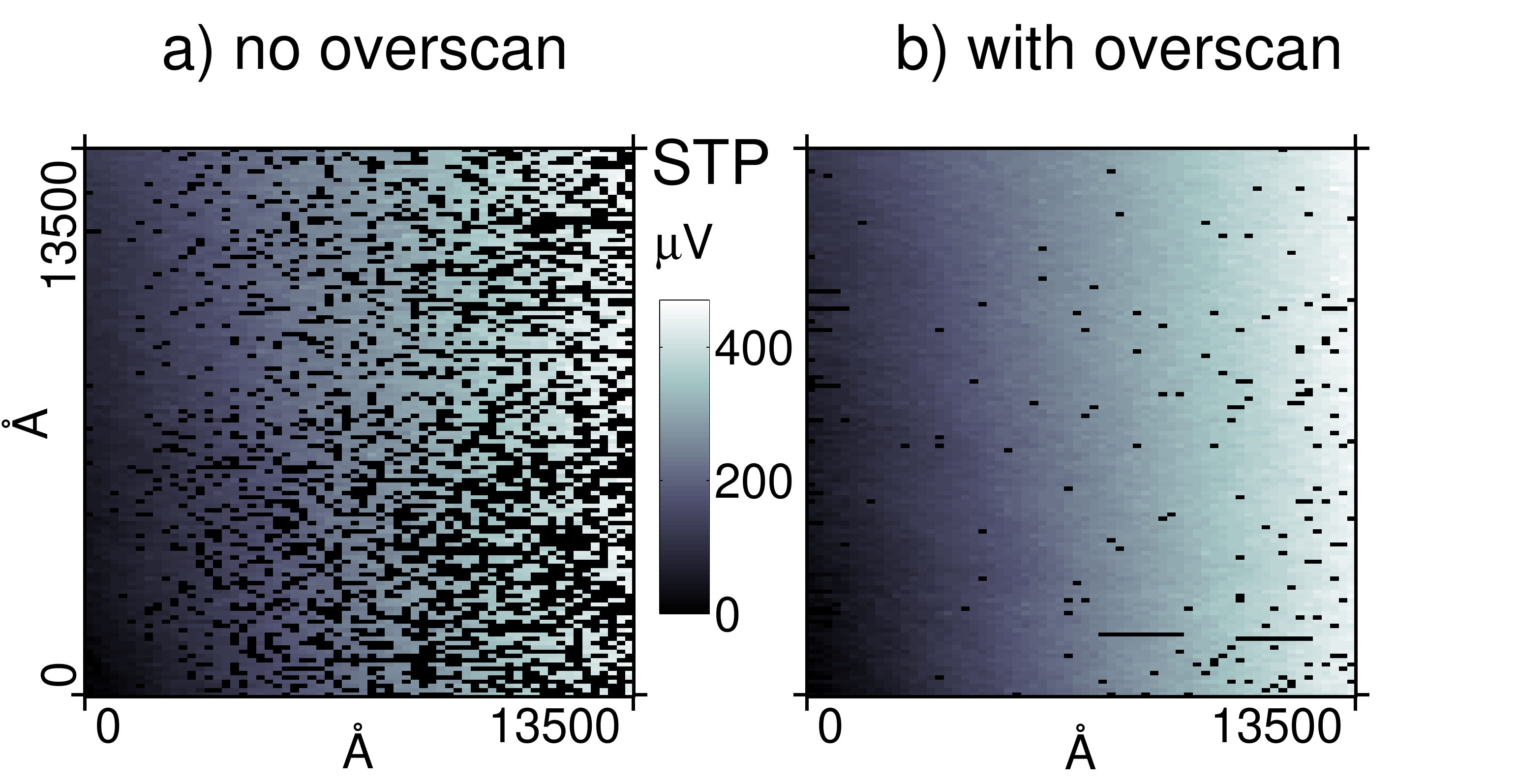}
\caption{Effect of piezo creep at large scan sizes. STP without and with point-by-point overscanning. Black pixels represent STP points rejected by the data acquisition algorithm due to piezo creep, which can be greatly reduced by the use of overscan, as can be seen by comparing a) and b). }
\label{fig:creeppic}
\end{figure}

As demonstrated in Section~\ref{basic}, the continuous zeroing of the tip-sample voltage by the potentiometric feedback allows scan range-independent low-noise STP performance. However, increasing scan sizes eventually leads to another problem, the resolution of which, while not critical, is desirable in a practical instrument. Figure~\ref{fig:creeppic} (left) shows a 1.35~$\mu$m square potentiometry scans acquired on a SrRuO$_{3}$ film with data acquired left to right on each scan line. The number of rejected points grows with increased scan sizes, and this problem has been observed for all types of samples studied. As described in Section~\ref{protocol}, such rejection occurs if the STP feedback loop fails to converge or a stable tunnel junction cannot be obtained. In the case discussed here, the rejection is due to the latter reason, with stable junctions being more difficult to achieve at large scan size due to piezo creep - the continued, uncommanded motion of a piezo after a commanded step. 

The resulting junction drift can be eliminated by longer delays after stepping, but the logarithmic time dependence of piezo creep makes this a painful approach. Using the fact that this hysteretic piezo effect is much weaker after the scanner motion is reversed (left side of the images), the problem can be addressed by introducing a turn-around into every step by overscanning -- moving the tip a distance that is longer than the needed step by a fixed numerical factor (overscan factor) , and then bringing it back to the desired position. The smaller number of skipped STP points in Figure~\ref{fig:creeppic} (right) , which shows STP data taken with an overscan factor of 7, but with the same STP acceptance criteria and measurement timing as before, confirms overscanning to be an efficient solution to the creep problem. 

\subsection{\label{resolution}Resolution for homogeneous transport}

The spatial resolution of an STP is not the same as the spatial resolution of the STM used. Rather, the resolution is given by how far the tip must be moved in a given electric field (E=$\rho$J in the continuum limit) to acquire a potential change larger than the noise.  Thus the resolution depends on both the sample resistivity ($\rho$), the current density (J), which is typically limited by sample heating, and by the noise of the STP itself. Mathematically, this can be written as  $\Delta$x=V$_{n}$/($\rho$J), where V$_{n}$ is the voltage noise of the measurement. To illustrate, we quantify the spatial resolution that can be reached with our STP in the case of thin films of SrRuO$_{3}$ and Au. The sample resistivities were 220~$\mu\Omega$-cm and 2.5~$\mu\Omega$-cm respectively, and average current densities of 10$^{4}$~A/cm$^{2}$ and 2.5$\times$10$^{5}$~A/cm$^{2}$ were used. SrRuO$_{3}$ data were obtained at junction resistance of 25~M$\Omega$ and measurement time constant of 1 sec; for the gold film these parameters were 6~M$\Omega$ and 2 sec.

\begin{figure}[h]
\includegraphics*[width=.48 \textwidth]{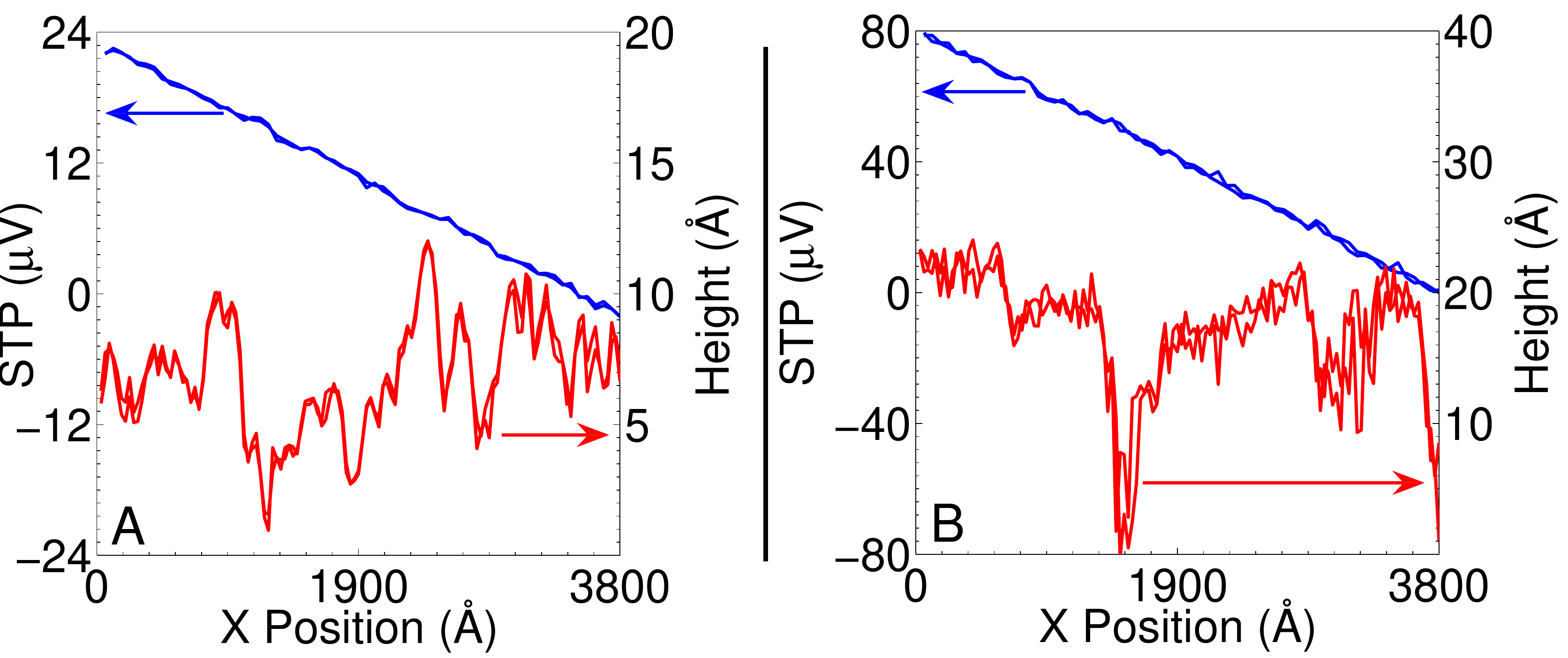}
\caption{Pairs of consecutive STP and topography linescans demonstrating effective spatial resolution on two representative materials: a) 30\AA-thick Au  (3800$\times$0~\AA~scan) and b) 558\AA-thick SrRuO$_{3}$ (3800$\times$3800~\AA~scan). The data demonstrate spatial resolution of the potential down to 18 and 24~\AA, respectively. In b), differences of topography scanlines are greater than 24~\AA~ due to the finite scan size in the y-direction (no effect on potential since gradient is along the x-direction).}
\label{fig:resolutionpic}
\end{figure}
Figure~\ref{fig:resolutionpic} shows consecutive topography and potentiometry linescans obtained on these samples at room temperature. Resolution can be extracted by comparing measured voltage noise to the voltage gradient measured over a known distance. The voltage noise for the  SrRuO$_{3}$ and gold measurements is 500~nV/$\sqrt{Hz}$ and 130~nV/$\sqrt{Hz}$, respectively, yielding 24~\AA~and 18~\AA~for resolution. The data demonstrate the ability of STP to resolve local voltages at nanometer length scales in these cases. Note that other scanning potentiometric methods, such as electrostatic force microscopy (EFM)\cite{efm} and conductive tip contact AFM-based potentiometry, may be capable of achieving good voltage sensitivity, but the effective large size of the probes used (long-range capacitive interaction for EFM or the poorly defined tip-sample contact region for contact techniques) limits their spatial resolution of potential. In contrast, for STP, the very local nature of the tunneling probe, with its effectively infinite spatial resolution, allows the resolution inferred from voltage sensitivity to determine the effective spatial resolution of the technique.

\subsection{\label{cryoop}Cryogenic operation}

All of the potentiometry measurements discussed in this work so far have been conducted at room temperature, albeit with careful temperature control. Happily, we have found that extending STP to lower temperatures does not introduce any new issues. As mentioned earlier, the system is pumped out into the 10$^{-5}$ Torr range in preparation for a cooldown. Once cryogenic temperatures have been established, the pressure drops into the 10$^{-7}$ Torr range, assisted partly by charcoal in the chamber. Cooling from room temperature to 4.2~K requires as little as 15 minutes. The flow rate needed to maintain this base temperature is about 4 liters of LHe per hour, allowing only 24 hours of uninterrupted measurement time for a 100 LHe storage dewar.  For operation at 100~K, the helium consumption rate drops by a factor of five. 

Down to the lowest temperatures, coarse motion was consistent in both the up and down directions, using approach waveforms as low as $\pm$75V. Using the capacitance between the tip and the sample as a measure of their separation allows the tip to be efficiently and rapidly moved to within a few dozen coarse steps from the sample prior to starting a fine approach. The total time from beginning a helium transfer with the sample withdrawn to imaging at 4.2~K can be as short as one hour.

\begin{figure}[h]

\includegraphics*[width=.48 \textwidth]{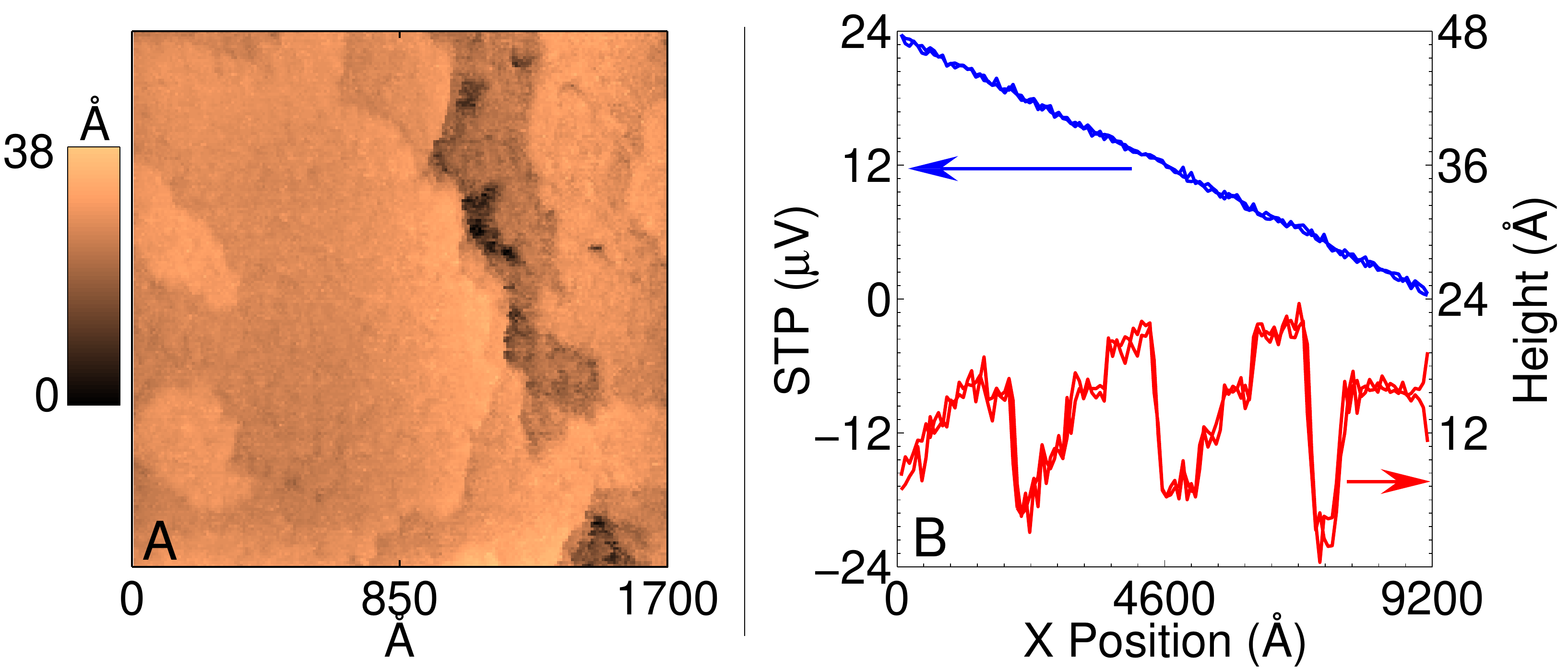}
\caption{Low temperature operation of STP. a) Imaging of SrRuO$_{3}$ at 4.2K. b) Consecutive topography and STP scanlines (from a different scan), demonstrating 65~\AA~spatial resolution of potential.  }
\label{fig:cooldownpic}
\end{figure}
	
STP measurements performed at 4.2~K on a SrRuO$_{3}$ film, shown in Figure~\ref{fig:cooldownpic}, show spatially homogeneous transport within the resolution of the instrument. Topography noise remains sub-angstrom, even in the presence of LHe flow and with no vibration isolation. As expected, with the tunnel junction at low temperature, the STP noise is dominated by the noise of the current amplifier. More specifically, for a 10~M$\Omega$ junction and a 1/4~Hz ENBW, the observed 180~nV rms noise is consistent with the 120~nV noise expected to be caused by the amplifier, as described in Section~\ref{basic}. With the measured gradient of 23~$\mu$V over 9100~\AA, the observed STP noise corresponds to local potential being spatially resolved to 65~\AA. 

The small, 30~$\Omega$  resistance of the sample prevented larger voltage gradients from being used for this measurement due to the limited compliance of the current source providing the gradient, but thinner or more resistive films should permit significantly higher local fields to be applied using existing electronics. The current source output driver can also be replaced by a higher current version in the future. Since the current amplifier noise scales linearly with the tunnel junction resistance (when referenced back to the sample) and current amplifiers with better noise characteristics are commercially available (for example, Keithley 428), STP measurements with spatial resolution of potentials approaching the atomic scale should be possible. 

\section{\label{rough}Rough samples and tip jumping}

\begin{figure}[h]
\includegraphics*[width=.5 \textwidth]{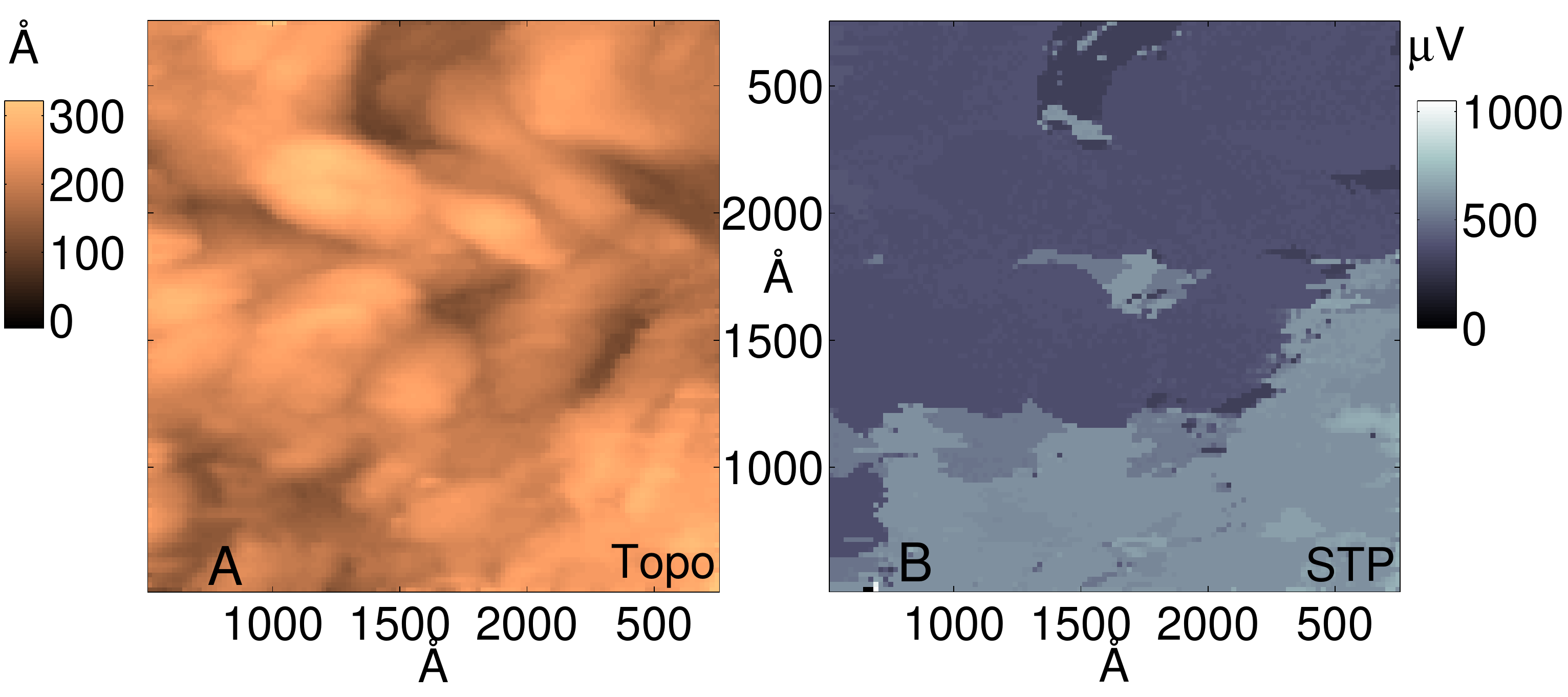}
\caption{Representative result on a rough surface: a) topography and b) potentiometry on a granular gold film with average applied gradient of 1.6~V/cm. Measured fields are up to 1000~V/cm with no clear relation to known gradient. The measurements are not reliable due to likely artifacts related to a finite size tip scanning a rough surface.}
\label{fig:oldstppic}
\end{figure}

All results discussed so far were obtained on smooth, uniform films and show voltage gradients that are consistent with known applied fields over long length scales. At most, small undulating features, the origin of which is not yet clear, can be discerned. Of course, not all materials of interest will be so smooth or uniform. As noted in the introduction, roughness can lead to artifacts associated with the finite size of the STM tips, and it is important to be able to distinguish between non-uniform potential maps due to such artifacts and those due to inhomogeneous transport properties. Indeed, our early measurements, taken on rougher surfaces presented a completely different picture than that seen on the smoother films. Figure~\ref{fig:oldstppic} shows typical topography and potentiometry obtained on a granular gold film biased to produce an average electric field of 1.6~V/cm. The data are qualitatively similar to most of the already referenced earlier results obtained using STP, with a notable exception of the work by Briner, et al\cite{feenstra}, in which great care was taken in preparing extremely smooth films.  As seen in the figure, measured local electric fields can be orders of magnitude higher than the applied field and are strongly correlated with topography. In addition, equipotential ``islands'' are evident, seemingly violating current conservation for a thin film (i.e., two-dimensional) sample.  

Although the very high local gradients imply spatial potential resolution essentially down to atomic scales, the data are impossible to interpret reliably, in the absence of any obvious relation to the known applied field. Less extreme behavior, in which the average gradient is evident but the abrupt features in potentiometry are still present, has also been observed. In all cases, the sharp steps in potentiometry scale with the applied gradient, indicating that they are related to the electrochemical potential created by the applied electric field.

It has previously been suggested\cite{kent,pelzkoch} that the discontinuous behavior illustrated in data such as these may be artifacts caused by imaging with a finite size tip on a rough surface.  The basic idea is that the exact location on the tip where tunneling occurs shifts from one point to another as the tip is scanned. Although this shift can be abrupt or gradual, we shall refer to all such effects as tip jumping, as they had been called historically. Although tip jumping is a general problem, affecting topographical imaging of any scanning probe, such artifacts can produce particularly dramatic features in STP. Obviously, in the presence of an electric field, a jump of the point of tunneling on the tip would lead to an abrupt change in the observed potential. Alternatively, in the event of sharp asperities on the surface effectively imaging a blunt tip, potentiometry images will show artificial potential plateaus. 

Despite the cogent warning by early investigators, subsequent studies failed to address systematically this possibility in the interpretation of their results, making definitive conclusions difficult. This historical situation remains confused and unresolved to this day. 

\begin{figure*}[htp]
\includegraphics*[width=1 \textwidth]{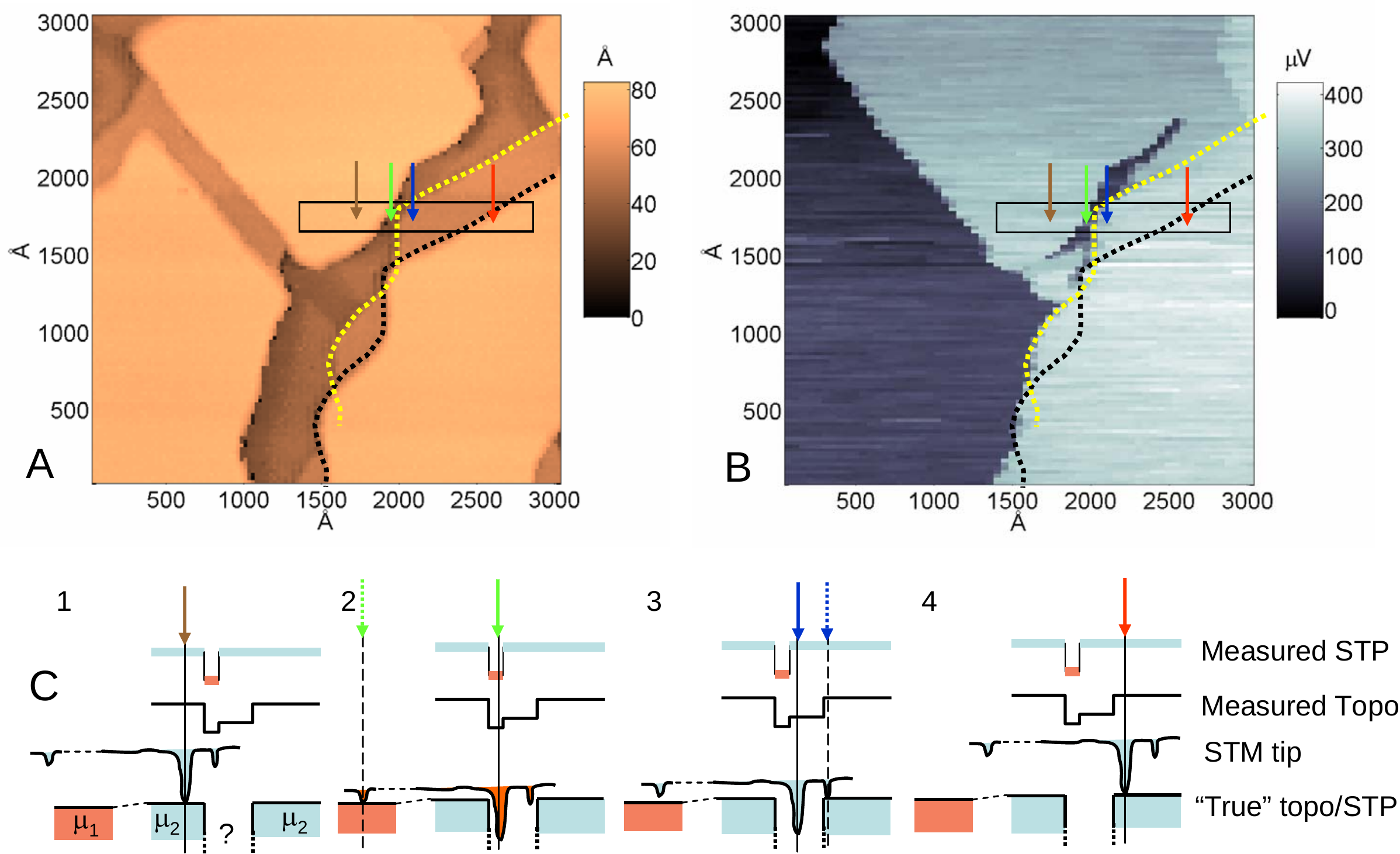}
\caption{Demonstration of the effects of tip jumping. a) Topography and b) potentiometry on an 86~\AA-thick~SrRuO$_{3}$ film in which deep trenches were intentionally introduced during growth. c) Likely scenario responsible for tip artifacts, as rectangular region is scanned left to right. Yellow and black dashed lines show obvious double image due to tip jumping in the topographic scan.  There is no immediately obvious correspondence with the potentiometric scan.  Panel c) shows how the potentiometry scan can be understood in terms of tip jumping. Solid colored arrows correspond to presumed tunnel location, indicated on images. Dashed colored arrows represent actual tunnel position, if different from presumed location.}
\label{fig:cartoonpic}
\end{figure*}

Since the effects of the finite size of STM tips are difficult to identify on a general rough surface and are often similar to the anticipated transport phenomena, we have examined the behavior of a model system (a very smooth film with intentional troughs) in order to clarify the nature and role of tip jumping artifacts in STP. Figure~\ref{fig:cartoonpic} shows topography/potentiometry data taken on a 86~\AA-thick SrRuO$_{3}$ film using a commercial STM tip\cite{tips}. The distinctly shaped topographic features easily enable identification of tip doubling, and the atomically flat plateaus in between the trenches allow the demonstration of both the correct functionality of STP and of the creation of STP artifacts by the interplay of the tip and surface geometry. It should be noted that some of our STP measurements on this material exhibit voltage noise above the tunnel junction noise level, which we attribute to resistance fluctuations in the material. This extra noise, which can be observed in a standard four-probe resistance measurement as well, does not affect the discussion of tip-jumping artifacts.

It is clear from the topographical image that the trenches produce tip-jumping-related image doubling (see the yellow and black dashed lines on the figure).   Clearly, as the tip was scanned over a trench, different protrusions on the tip contact the sample, resulting in features being doubled. The lateral separation of the double image, and concomitantly of the protrusions responsible for tip-doubling, is 400~\AA~in this case.  Note, however, that the height difference is only about 15~\AA. Despite such a ``flat'' tip aspect ratio, the imaging remains good, since the tip jumping is being caused by tip geometry as opposed to some dynamic events. Therefore, image quality cannot be used to dismiss the possibility of multiple/flat tips, as previously suggested\cite{grevin}. Also, a close look at the image reveals the fact that the bottom of the trenches is never actually imaged, with no information obtained about the topography/potential of such regions. True imaging is obtained only when the lowest point of the tip is involved in the tunneling, and can be discerned on materials with very distinct topographical features.

The tip-related effects responsible for the artifacts visible in the topographic image have a dramatic impact on the STP data. The deep trenches seen in Figure~\ref{fig:cartoonpic} dominate the sheet resistance of the sample, with most of the voltage drop occurring across the trenches and no measurable voltage gradients evident on the plateaus at the indicated level of voltage sensitivity.  Some trenches do not have an associated voltage drop, presumably since there is no current flowing across them as a result of some torturous current path.  However, while the observed voltage drops across the trenches are real, many of the exact details are artifacts caused by the finite tip size and serve to illustrate this effect.

At the bottom of Figure~\ref{fig:cartoonpic} we illustrate schematically a possible scenario for how the observed topographic and potentiometric images could arise due to tip jumping. Initially (frame 1), with the lowest protrusion of the tip above a plateau, the presumed and actual tunneling locations overlap, yielding correct STP and Z data. As the scan continues past the trench edge, different protrusions reach the surface sequentially, causing actual tunneling locations to jump laterally. This leads to a number of artifacts for both topography and potentiometry, ranging from instantaneous voltage steps to creating the voltage islands. Some of the artifacts lead to image doubling, and are obvious in topographic images; others cannot be detected as easily. Notably, the area of the sample to which the tip location switches to create the low voltage island is not identifiable from topography, either because the tip ends up on a featureless region of one of the plateaus or outside of the presumed image region altogether; thus tip-jumping artifacts can occur without being identifiable in the topographic image. Eventually (frame 4), the ``true'' tip again becomes the actual tunneling location, again yielding correct data. Essentially, the data obtained on the plateaus is always valid, while information obtained inside the trench may not be. We believe that the observations presented here confirm the existence of tip-related artifacts, which are the likely explanation for many earlier results obtained by scanning potentiometry on the rougher samples. The large lateral extent of the tips can cause not just the artificial sharpening of potential steps but also create a completely false measured landscape of local potential.

\begin{figure*}[htp]
\includegraphics*[angle=90,width=1 \textwidth ]{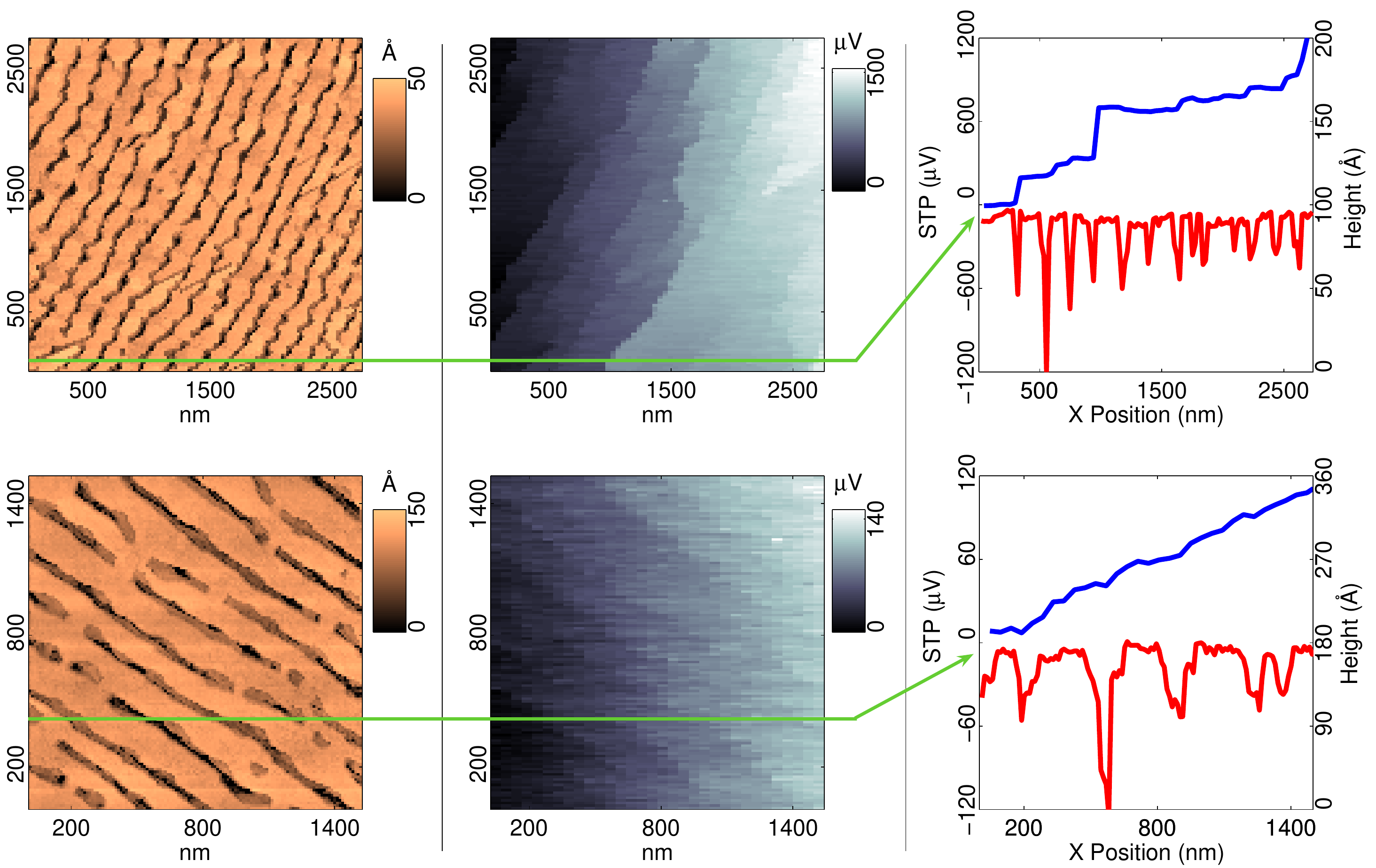}
\caption{Transport across deep trenches in 88~\AA- (top) and 300\AA-thick (bottom) SrRuO$_{3}$ films. Left: topography, center: STP, right: topography and potentiometry linescans (location indicated by the green lines on images). STP clearly shows the greater impact of trenches on transport in thinner films.} 
\label{fig:trenchespic}
\end{figure*}

Although, as we have shown here, artifacts of the type described by Pelz and Koch are possible in STP, they are by no means fatal.  Preferably, scanning potentiometry, like many other scanning techniques, should be performed on flat samples, for studies of properties not related to surface morphology. But even if large topographic features are unavoidable, many of the phenomena that can be studied will be switchable or tunable, for example by an applied field or temperature changes, enabling their separation from topographic artifacts. 

Finally, as in the extreme case of studying the effects of surface microstructure on the transport, the tip-related artifact problem can still be managed to produce meaningful results. As we have demonstrated, with good topographic imaging and distinct surface features (as opposed to ``rolling hills'' morphology on many granular materials), it is possible to distinguish the regions in which topographic and voltage data can be trusted from those which are susceptible to artifacts. For the SrRuO$_{3}$ films discussed here, data from plateaus is always valid and can provide information on how the trenches are affecting the transport. 

Figure~\ref{fig:trenchespic} shows topography and potentiometry on two SrRuO$_{3}$ films of different thicknesses, 86~\AA~and 300~\AA. The sheet resistance (from four-probe resistance measurements) of the thinner sample is twenty times higher than that of the thicker one, which is not surprising since the resistance of the thinner sample is more affected by the deep trenches. This is confirmed by STP, which shows all voltage drops occurring across the trenches for the 86~\AA~film, whereas for the 300~\AA~sample the resistance is shared evenly with the flat regions. It can also be seen that there is no obvious correspondence between the trench depth, as measured by the STM and trench resistance, as measured by STP. This is due to the fact that, as discussed above, tip geometry prevents correct imaging of the trenches, leading to erroneous depth information as the trench effectively images the tip. However, the partial but valid information obtained by potentiometry on the flat regions, reveals actual trench depth more accurately. Thus, STP can sometimes be used to obtain topographic information that one cannot get with a scanned topographic probe alone. 

\section{\label{improve}Possible improvements}

With the measurement pushed to the fundamental noise floor, few performance improvements are possible beyond those already mentioned above. Therefore, most meaningful enhancements would be those that expand the versatility of the instrument. With tip-related artifacts placing some constraints on the types of surfaces that can be easily studied with STP, a reliable technique for obtaining high aspect ratio tips would be most welcome. Mounting metallic carbon nanotubes on the tip, as originally proposed by Dai, et al\cite{dai}, would greatly reduce or eliminate tip-jumping artifacts on any surface. 	

A promising application of STP would be the study of local transport in artificial nanostructures, with obvious basic and applied science uses. Since an STM needs a conducting surface to operate, the ability to produce interesting devices to study with the current instrument is limited. Functionality can be expanded by combining AFM and STM capability, as discussed, for example, in the work by Giessibl\cite{giessibl}, in which an STM tip is mounted on a quartz tuning fork. The microscope can then be used in a non-contact AFM mode over insulating areas and switched to the STP mode once the conducting region is located. 

A final addition would be the incorporation of standard scanning spectroscopy capability to the system, to add LDOS information to topography and potentiometry. Since all the hardware for this task is already in place, only some extra software would be needed; however, due to the increased vibrational requirements of STS, more sophisticated vibration isolation solutions may be required.

\begin{acknowledgments}
We gratefully acknowledge several contributions: S.H. Pan gave useful advise on microscope design. High voltage amplifiers used for scanning we duplicated from a system designed by Alan Fang. SrRuO$_3$ and smooth Au films were grown by Wolter Siemons and Gertjan Koster. This work was supported by the Air Force Office of Scientific Research through a Multi-University Research Initiative (MURI), and in part by Center for Probing the Nanoscale, an NSF NSEC, NSF Grant No. PHY-0425897.
\end{acknowledgments}


\newpage


\begin{references}

\bibitem{muralt}
P. Muralt and D. W. Pohl, Applied Physics Letters \textbf{48}, 514 (1986)

\bibitem{contactafm}
T. Trenkler, P. De Wolf, W. Vandervorst, and L. Hellemans, J. Vac. Sci. Technol. B \textbf{16}, 367 (1998)

\bibitem{feenstra}
B. G. Briner, R. M. Feenstra, T. P. Chin, and J. M. Woodall, Phys. Rev. B \textbf{54}, R5283 (1996)

\bibitem{kirtley}
J. R. Kirtley, S. Washburn, and M. J. Brady, Phys. Rev. Lett. \textbf{60}, 1546 (1988)

\bibitem{kent}
A. D. Kent, I. Maggio-Aprile, Ph. Niedermann, Ch. Renner, and \O. Fischer, J. Vac. Sci. Technol. A \textbf{8}, 459 (1990)
%  Scanning tunneling potentiometry studies of Y1Ba2Cu3O7-x and gold thin films
%    A. D. Kent, I. Maggio-Aprile, Ph. Niedermann, Ch. Renner, and Ø. Fischer 

\bibitem{besold}
J. Besold, R. Kunze, and N. Matz, J. Vac. Sci. Technol. B \textbf{12}, 1764 (1994)
%  Electromigration kinetics of gold on a carbon thin film surface studied by scanning tunneling microscopy and scanning tunneling potentiometry

\bibitem{schneider}
M. A. Schneider, M. Wenderoth, A. J. Heinrich, M. A. Rosentreter, and R. G. Ulbrich, Applied Physics Letters \textbf{69}, 1327 (1996)
%  Current transport through single grain boundaries: A scanning tunneling potentiometry study

\bibitem{ramaswamy}
Geetha Ramaswamy and A. K. Raychaudhuri, Applied Physics Letters \textbf{75}, 1982 (1999)
%  Field and potential around local scatterers in thin metal films studied by scanning tunneling potentiometry
%Appl. Phys. Lett. -- September 27, 1999 -- Volume 75, Issue 13, pp. 1982-1984
%Geetha Ramaswamy, A. K. Raychaudhuri

\bibitem{grevin}
B. Gr\'{e}vin, I. Maggio-Aprile, A. Bentzen, L. Ranno, A. Llobet, and \O. Fischer, Phys. Rev. B \textbf{62}, 8596 (2000)
%Local electronic transport in La0.7Sr0.3MnO3 thin films studied by scanning tunneling potentiometry
%Phys. Rev. B 62, 8596 - 8599 (2000)

\bibitem{pelzlownoise}
P. Pelz and R. H. Koch, Rev. Sci. Instrum. \textbf{60}, 301 (1989)
%  Extremely low-noise potentiometry with a scanning tunneling microscope

\bibitem{pelzkoch}
P. Pelz and R. H. Koch, Phys. Rev. B \textbf{41}, 1212 (1990)
%  Tip-related artifacts in scanning tunneling potentiometry, JP Pelz, RH Koch  

\bibitem{thesis}
A more detailed description of microscope and system design can be found in: Michael Rozler, Ph.D. thesis, Stanford University, 2008

\bibitem{pan}
S. H. Pan, E. W. Hudson, and J. C. Davis, Rev. Sci. Instrum. \textbf{70}, 1459 (1999)
%  3He refrigerator based very low temperature scanning tunneling microscope

\bibitem{hug}
Hans J. Hug, B. Stiefel, P. J. A. van Schendel, A. Moser, S. Martin, and H.-J. G\"untherodt, Rev. Sci. Instrum. \textbf{70}, 3625 (1996)
%  A low temperature ultrahigh vaccum scanning force microscope

\bibitem{desert}
Desert Cryogenics, currently division of Lakeshore Cryotronics.

\bibitem{newport}
Four CM-225 isolators, Newport Corporation.

\bibitem{apex}
Apex Microtechnology Application Note 13 - Voltage to Current Conversion.

\bibitem{efm}
Y. Martin, D. W. Abraham, and H. K. Wickramasinghe, Applied Physics Letters \textbf{52}, 1103 (1988)

\bibitem{tips}
Veeco Probes model PT STM tip.

\bibitem{dai}
H. Dai, J. H. Hafner, A. G. Rinzler, D. T. Colbert, and R. E. Smalley, Nature \textbf{384}, 147 (1996)
%  Nanotubes as nanoprobes in scanning probe microscopy

\bibitem{giessibl}
Franz J. Giessibl, Applied Physics Letters \textbf{76}, 1470 (2000)
%  Atomic resolution on Si.111.-.7Ã7. by noncontact atomic force
%microscopy with a force sensor based on a quartz tuning fork, 

\end{references}
\end{document}